# Metal (boro-) hydrides for high energy density storage and relevant emerging technologies

L.J. Bannenberg,[a] M. Heere,[b,c] H. Benzidi,[d] J. Montero,[e] E.M. Dematteis,[e,f] S. Suwarno,[g] T. Jaroń,[h,*] M. Winny,[h,i] P. A. Orłowski,[h,j] W. Wegner,[h,i] A. Starobrat,[h,i] K.J. Fijałkowski,[h] W. Grochala,[h] Z. Qian,[k] J.-P. Bonnet,[l] I. Nuta,[m] W. Lohstroh,[c] C. Zlotea,[e] O. Mounkachi,[d] F. Cuevas,[e] C. Chatillon,[m] M. Latroche,[e] M. Fichtner,[n,o] M. Baricco,[f] and B.C. Hauback,[p] A. El Kharbachi [n,*]

[a] Faculty of Applied Sciences, Delft University of Technology, Mekelweg 15, 2629 JB Delft, The Netherlands

[b] Institute for Applied Materials—Energy Storage Systems (IAM–ESS), Karlsruhe Institute of Technology (KIT), 76344 Eggenstein, Germany

[c] Heinz Maier-Leibnitz Zentrum, Technische Universität München, 85748 Garching, Germany

[d] Laboratory of Condensed Matter and Interdisciplinary Sciences (LaMCScI), Associated to CNRST (URAC 12), Physics Department, Faculty of Sciences, Mohammed V University, Rabat, Morocco

[e] Univ. Paris Est Creteil, CNRS, ICMPE, UMR7182, F-94320, Thiais, France

[f] Department of Chemistry and NIS, University of Turin, Via P.Giuria, 9, I-10125, Torino, Italy

[g] Department of Mechanical Engineering, Institut Teknologi Sepuluh Nopember (ITS), Kampus ITS Keputih Surabaya, Indonesia 6011

[h] Centre of New Technologies, University of Warsaw, Banacha 2C, 02-097 Warsaw, Poland

[i] College of Inter-Faculty Individual Studies in Mathematics and Natural Sciences, University of Warsaw, Banacha 2C, 02-097 Warsaw, Poland

[j] Faculty of Chemistry, University of Warsaw, Pasteura 1, 02-093 Warsaw, Poland

[k] Key Laboratory of Liquid-Solid Structural Evolution and Processing of Materials (Ministry of Education), School of Materials Science and Engineering, Shandong University, Jinan 250061, China

[l] Laboratoire de Réactivité et Chimie des Solides (CNRS UMR 7314), Université de Picardie Jules Verne, 33 Rue Saint Leu, 80039 Amiens Cedex, France

[m] Univ. Grenoble Alpes, CNRS, Grenoble INP*, SIMaP, 38000 Grenoble, France

[n] Helmholtz Institute Ulm (HIU) Electrochemical Energy Storage, Helmholtzstr. 11, 89081 Ulm, Germany

[o] Institute of Nanotechnology, Karlsruhe Institute of Technology (KIT), P.O. Box 3640, 76021 Karlsruhe, Germany

[p] Institute for Energy Technology, P.O. Box 40, NO-2027 Kjeller, Norway

Corresponding Authors: t.jaron@cent.uw.edu.pl and kharbachi@kit.edu

# Abstract

The current energy transition imposes a rapid implementation of energy storage systems with high energy density and eminent regeneration and cycling efficiency. Metal hydrides are potential candidates for generalized energy storage, when coupled with fuel cell units and/or batteries. An overview of ongoing research is reported and discussed in this review work on the light of application as hydrogen and heat storage matrices, as well as thin films for hydrogen optical sensors. These include a selection of single-metal hydrides, Ti-V(Fe) based intermetallics, multi-principal element alloys (high-entropy alloys), and a series of novel synthetically accessible metal borohydrides. Metal hydride materials can be as well of important usefulness for MH-based electrodes with high capacity (e.g. $MgH_2$ ~ 2000 mAh $g^{-1}$) and solid-state electrolytes displaying high ionic conductivity suitable, respectively, for Li-ion and Li/Mg battery technologies. To boost further research and development directions some characterization techniques dedicated to the study of M-H interactions, their equilibrium reactions, and additional quantification of hydrogen concentration in thin film and bulk hydrides are presented at the end of this manuscript.

# 1. Introduction

It was already in the nineteenth century when British chemist Thomas Graham discovered that palladium can accommodate vast amounts of hydrogen 'being eight to nine hundred times its volume in hydrogen gas' [1]. In the years following this pioneering study, it was found that metal hydrides may accommodate non-stoichiometric amounts of hydrogen up to a hydrogen-to-metal ratio of $H/M = 3$ for some transition metals, and even higher at $P$ of hundreds GPa. Consequently, some metal hydrides have remarkably high atomic hydrogen densities (e.g. $N_H = 14.4 \cdot 10^{28}$ m$^{-3}$ for VH$_2$) that well exceed the density of liquid hydrogen ($N_H = 4.2 \cdot 10^{28}$ m$^{-3}$) [2]. The hydrogenation of metals may dramatically alter the electronic, optical and magnetic properties of materials, inducing metal-to-insulator changes or changing paramagnets into superconductors at low temperature [2,3].

Starting from the 1970s, metal hydrides were considered as an ideal candidate for hydrogen storage because of their high atomic hydrogen density. Hydrogen – a clean and sustainable energy carrier with the highest energy density per mass of fuel – will likely play a paramount role in the transition to a carbon neutral economy. Hydrogen can be particularly attractive for mobile applications, for instance powering heavy-duty vehicles as well as for stationary energy storage where it may be used to match the relatively constant energy demand with the fluctuating output of renewable energy sources (solar, wind, hydropower, etc.). Yet, under ambient conditions, hydrogen gas has a low energy density per volume unit, and viable hydrogen storage is therefore key for its successful implementation. Compared to other methods of hydrogen storage, storage in metal hydrides, apart from the high volumetric density, has the particular advantage that no cryogenic cooling is required (as for liquid hydrogen storage) and that is intrinsically safer than high-pressure storage where potential leaks form a safety hazard. In particular, research focused on lightweight metal hydrides such as magnesium-based and, so called, "complex metal hydrides" that also feature relatively high gravimetric hydrogen densities [4-8]. The latter term denotes multinary ionic compounds composed of cations and polyatomic anions containing hydrogen, in principle covalently bound to the heavier atom, *e.g.* NaBH$_4$, Li$_2$B$_{12}$H$_{12}$ or LiAlH$_4$. The term "complex hydrides" has even been stretched to the compounds containing hydrogen in non-hydridic but rather protonic form, like amides, or imides, *e.g.* NaNH$_2$ or Li$_2$NH.

Although most metal hydrides exhibit a high energy density, their applicability is in general limited both by thermodynamic and kinetic factors. Hydrogen forms relatively strong bonds with most metals, implying reasonably large negative values for the enthalpy and entropy of formation (e.g. $\Delta H \approx -74$ kJ mol$_{H2}^{-1}$ and $\Delta S \approx -135$ J K$^{-1}$ mol$_{H2}^{-1}$ for MgH$_2$ [9,10]). As a consequence, hydrogenation readily occurs exothermically under mild conditions, while the endothermic desorption requires high temperatures and the supply of a certain amount of heat. In addition, although the diffusion of hydrogen in metals is in general much faster than the diffusion of other compounds, slow hydrogen (de)sorption as well as poor cyclability remain a problem in most materials [7,8,11].

Especially in the 1990s and early 2000s, various strategies have been employed to overcome these challenges including alloying and nanostructuring [12,13]. Although these strategies considerably improve the storage properties of metal hydrides, none of the investigated materials simultaneously meet at the moment the requirements on desorption temperature and pressure, gravimetric and volumetric energy density, cyclability, (de)sorption kinetics and economic feasibility. In other words, continuous efforts are still required to obtain breakthrough materials with high storage efficiency and cost-effective processing. Although advances have been made in the last decades, the prospective advantages of hydrogen storage in metal hydrides faded partly away owing to the rapid development of alternative technologies as Li-ion batteries and gaseous hydrogen storage in light-weight high pressure vessels.

Meanwhile, new applications of metal hydrides have emerged. Metal hydrides based compressors are already commercially available [14], and metal hydrides are also present in many commercial batteries (e.g. Ni*M*H [15,16]). Owing to their light-weight, low voltage and small hysteresis compared to oxide anodes, metal hydrides are now intensively studied for their applications as high energy density electrodes for LIBs with carbonate-based liquid electrolytes [17-19]. In addition, complex metal hydrides are studied for their application as solid-state electrolytes in next-generation all-solid-state batteries [20-23]. In these batteries, the highly flammable liquid electrolyte is replaced by a solid one, making these batteries intrinsically safer and potentially increasing the energy density. Furthermore, metal hydride based hydrogen sensors are now considered as a competitive way to reliably and efficiently sense hydrogen over extensive hydrogen pressure ranges [24-28]. These sensors utilize the profound change of the electronic and/or optical properties of most metal hydrides when (partially) hydrogenated as a result of an exposure to a hydrogen environment.

The purpose of this review is to provide an overview of various metal hydrides with a particular focus on applications in a green economy. In Section 2 we discuss single-metal hydrides, multi-metallic systems and a series of metal borohydrides with focus on synthesis, structural and hydrogen storage properties. While Section 3 deals with the application of metal hydrides in hydrogen sensors, Section 4 discusses the incorporation of metal hydrides in rechargeable batteries, including conversion-type electrodes, together with lithium and magnesium-based solid-state electrolytes. Section 5 describes experimental techniques relevant for the study of metal hydrides. In Section 6 we summarize the main recent results and provide an outlook on the application of metal (boro-)hydrides. For guidance throughout this review report abbreviations and acronyms are listed below (Table 1).

Table 1. The acronyms and abbreviations used across this review.

| | |
|---|---|
| AIMD | *ab initio* molecular dynamics |
| ASSB | all-solid-state battery |
| *bcc* | body-centered cubic |
| *bct* | body-centered tetragonal |
| CVD | chemical vapor deposition |
| DFT | density functional theory |
| DMC | dimethyl carbonate |
| DME | dimethoxyethane |
| DMS | dimethyl sulfide |
| DSC | differential scanning calorimetry |
| EC | ethylene carbonate |
| EIS | electrochemical impedance spectroscopy |
| en | ethylenediamine |
| *fcc* | face-centered cubic |
| FTIR | Fourier-transform infrared spectroscopy |
| *hcp* | hexagonal close packed |
| HEA | high-entropy alloy |
| *h*-LiBH$_4$ | hexagonal LiBH$_4$ |
| HP | high-pressure |
| HPT | high-pressure torsion |
| IDT | inter-digitated transducer |
| KE | Knudsen effusion |
| LIB | Li-ion battery |
| (L)SPR | (localized) surface plasmon resonance |
| MEMS | microelectromechanical-system |
| MH | metal hydride |
| MOF | metal-organic framework |
| MPEA | multi-principal element alloy |
| MS | mass spectrometry |
| NMR | nuclear magnetic resonance |
| NRA | nuclear reaction analysis |
| *o*-LiBH$_4$ | orthorhombic LiBH$_4$ |
| *P* | pressure |
| PCT | pressure-composition isotherm |
| PDF | pair distribution function |
| PEM | proton-exchange membrane |
| PMMA | poly(methylmethacrylate) |
| PTFE | polytetrafluoroethylene |
| PTI | pressure-optical transmission isotherm |
| QENS | quasi-elastic neutron scattering |
| *RE* | rare earth |
| RMB | rechargeable magnesium battery |
| SE | solid electrolyte |
| SHE | standard hydrogen electrode |
| SLD | scattering length density |
| SQUID | superconducting quantum interference device |
| SR-XRD | synchrotron radiation - powder X-ray diffraction |
| *T* | temperature |
| TEM | transmission electron microscopy |
| THF | tetrahydrofuran |
| TOFTOF | high-resolution time-of-flight spectrometer |
| UHV | ultra-high vacuum |
| VEC | valence electron concentration |
| WCA | weakly coordinating anions |
| XPS | X-ray photoelectron spectroscopy |

## 2. Metal hydrides for solid-state hydrogen storage

### 2.1. Single-metal hydrides

Considering stable (non-radioactive) elements, all of the main group and many transition metals form binary hydrides, $MH_n$, either under mild ($P$, $T$) or harsh high pressure (HP) conditions [29]. Here we will briefly refer to this group of hydrides in the context of hydrogen storage. Among such simple hydrides, only those containing the most lightweight elements are potentially able to fulfil the requirement of a sufficiently high gravimetric capacity for onboard hydrogen storage for light-duty fuel cell vehicles of 6.5 wt% [30], while this parameter remains not as severely limiting for the multinary hydrides (Fig. 1). At the same time, significant volumetric density of hydrogen may be achieved even for heavier binary hydrides – while lightweight LiH of 12.6 wt% H stores 98.3 kg m$^{-3}$. Much heavier $BaH_2$ containing only 1.45 wt% H has a hydrogen density of 60.5 kg m$^{-3}$, which still fulfils the U.S. Department of Energy (DOE) target (updated May 2017) of > kg m$^{-3}$ [30]. Despite limited options for gravimetrically efficient hydrogen storage, single-metal hydrides are still an appealing class of materials for energy storage in potential applications other than light-duty vehicles. This is related to their simplicity, usually combined with emission of pure hydrogen in a simple, one- or two-step thermal decomposition process, according to eq. (1):

$$MH_n \rightarrow M + n/2 H_2 \uparrow \qquad (1)$$

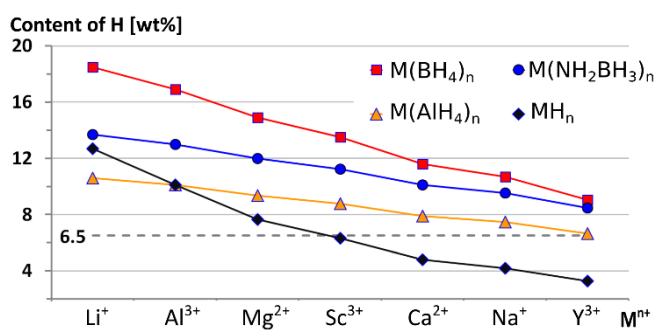

**Figure 1.**

Comparison of the nominal gravimetric content of hydrogen in simple $MH_n$ and selected multinary (complex) hydrides ($M(BH_4)_n$, $M(AlH_4)_n$) and ($M(NH_2BH_3)_n$) for various metals, M.

Single-metal hydrides are traditionally classified into three groups: ionic, covalent and interstitial. This division works fine mostly for a few typical examples in each category, however, with numerous compounds of intermediate properties [31]. The typical ionic hydrides (also known as "saline hydrides") are formed by most of the s-block metals: Li – Cs, Mg – Ba, however, already $MgH_2$ exhibits a partial covalent character of its metal-hydrogen bonding [32]. These non-volatile and insulating compounds share densely-packed crystal structures with metal halides, fulfilling the known hydride-fluoride analogy [33]. Accordingly, the NaCl-type is adopted by MH, M = Li – Cs. Similarly to $MgF_2$, $MgH_2$ crystallizes in rutile ($TiO_2$) structure, while $MH_2$, M = Ca – Ba, are isostructural with $PbCl_2$, common with fluorite ($CaF_2$) above ~10 GPa [34,35]. Also the lanthanide trihydrides, $LnH_3$, as well as $EuH_2$ and $YbH_2$ can be classified as ionic, unlike metallic $LnH_2$ and the non-stoichiometric compounds [36]. Beryllium and the p-block metals form hydrides of significantly covalent character, which adopt 0D molecular crystals, such as volatile $SnH_4$ [37], or polymeric 3D networks, *e.g.* via corner sharing of corresponding tetrahedra or octahedra like $BeH_2$ and $AlH_3$, respectively [38,39]. The binary hydrides of transition metals, some hydrides of lanthanides of a formula $LnH_{n<3}$, and various actinide hydrides are classified as interstitial hydrides [40-42]. These compounds are often non-stoichiometric across a broad composition range, they show metallic cluster and usually conduct electricity well; conductivity is strongly affected by the molar content of hydrogen which influences electronic occupancy of the conduction band. Many hydrides belonging to this group form by incorporation of hydrogen atoms into suitable interstices (octahedral *O*, or tetrahedral *T*, vacancies) in the metallic lattice. However, this process usually occurs with substantial volume change (often 10–20% per hydrogen atom) and is often accompanied by hydrogen-induced phase transitions [43,44]. Such hydrides often show significant mobility of hydrogen atoms, mainly at elevated temperatures. This feature is especially noticeable for palladium and allows for "filtration" of hydrogen through membranes based on Pd, or its alloys, resulting in 99.99999% hydrogen purity [45,46]. Interestingly, palladium is able to reversibly absorb ~400 times its own volume of hydrogen at ambient conditions, while at lower temperature and elevated pressure this value raises almost to 1000. Remarkably, absorption of $H_2$ is connected with drop of the molar volume of the metal in this case. Unfortunately, $PdH_{0.6}$, with a mere hydrogen capacity of 0.57 wt% H, is a typical stoichiometry which can be reached for palladium-hydrogen system near ambient conditions, and there is no evidence for H:Pd atomic ratio larger than 1 even up to 100 GPa [47]. These facts, in addition to the high price of palladium, exclude $PdH_n$ from broader applications in hydrogen storage; however, due to its excellent reversibility, palladium tritide $PdT_x$, remains a common tritium reservoir in nuclear facilities [48,49].

Besides the systems' hydrogen capacity, the temperature of the thermal decomposition ($T_{dec}$) is connected with the release of hydrogen and, therefore, one of the key practical parameters related to hydrogen-storage materials. This parameter has been thoroughly analyzed for a number of hydrides. For most of the $MH_n$ compounds the values of $T_{dec}$ well correlate with their thermodynamic parameters, especially with the free energy ($\Delta G°$) of the reaction, the standard redox potential ($E°$) of the $M^{n+}/M^0$ redox couple and with the standard enthalpy of $MH_n$ formation ($\Delta H°_f$) [31]. It appears that $T_{dec}$ remains high (330 – 720 °C) for the highly electropositive elements with $E° < -2.3$ V. $T_{dec}$ adopts moderate temperature values (0 – 250 °C) for moderately electropositive metals and metalloids ($E°$ between –2.0 and –0.6 V), while the electronegative metals and most of the semimetals studied ($E°$ between –0.6 and +0.85 V) are rather unstable ($-125 < T_{dec} < -15$ °C). While this relation remains monotonic for most of the binary hydrides studied, there are a few exceptions, which are either more or less stable than the prevailing majority of $MH_n$ compounds. Some of these discrepancies can be explained by possible excessive kinetic stability, like for CuH, which should remain in equilibrium with the gaseous hydrogen only above an immense pressure of 8.5 GPa, while it shows a half-life of *ca.* 30 h at ambient conditions [50-52].

Due to the above-mentioned restrictions on the gravimetric energy density, LiH, $MgH_2$ and $AlH_3$ are the binary systems that have been predominantly studied as potential hydrogen storage materials. Although some tuning of the conditions under which hydrogen can be desorbed appeared possible, still non-satisfactory parameters have been achieved. The first of these compounds, LiH, is the most thermally stable among these metal hydrides ($T_{dec} \approx 720\ ^oC$, decomposition is preceded by melting at *ca.* 689 $^oC$). This temperature has been significantly decreased by doping with Si (Li : Si = 2.35 : 1), which allows for reversible storage of *ca.* 5 wt% H, released during heating to 490 $^oC$ [53]. In the case of $MgH_2$ slow kinetics of de-/re-hydrogenation and high formation enthalpy (–75.2 kJ $mol^{-1}$) are the main problems [31]. Due to such parameters, heating to >400 $^oC$ is necessary for reversible hydrogen storage. Several methods for improvement have been tested for this parent system like the decreasing of the grain size and introduction of defects *via* high-energy ball-milling [54], forming nano-fibers utilizing CVD [55], incorporation of $MgH_2$ into the carbon aerogel nanoscaffold [56], destabilization via ion irradiation [57,58], or using various catalysts [53,59,60]. While some of them led to improved kinetics and lowered temperature of hydrogen release, these achievements are still far from meeting the targets for on-board hydrogen storage [61]. In contrast to the two previous hydrides, $AlH_3$ is a metastable compound, strongly stabilized by significant barrier of decomposition and due to easy surface passivation by traces of oxygen. Although its room-temperature equilibrium pressure of $H_2$ is close to 50 GPa ($\Delta G^o_f = +48.5 \pm 0.4$ kJ $mol^{-1}$, $\Delta H^o_f = -9.9 \pm 0.4$ kJ $mol^{-1}$), its thermal decomposition occurs around 150 $^oC$ [61], or around 100 $^oC$ for the freshly prepared compound [62,63]. Therefore, despite fair kinetics of hydrogen release, the reversibility of this system in the acceptable pressure range remains the sole unsolved problem [64].

It is worth to mention that preparation of so called "reactive hydride composites" is one of the tuning options tested for the mixtures of simple and complex metal hydrides [65-68]. Such systems are selected in order to facilitate thermally-induced chemical reactions between their components to change the original (*i.e.* that for the parent chemical compounds) hydrogen release path and lower the overall reaction enthalpy. For example the $MgH_2/2LiBH_4$ composite [69] decomposes according to the following scheme:

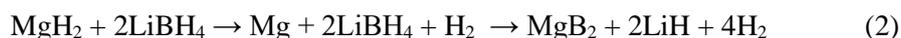

$$MgH_2 + 2LiBH_4 \rightarrow Mg + 2LiBH_4 + H_2 \rightarrow MgB_2 + 2LiH + 4H_2 \qquad (2)$$

This system shows dramatic improvement of kinetics of hydrogen absorption, which is observed already at 250 $^oC$ under 5 MPa hydrogen, while 600 $^oC$ and 35 MPa $H_2$ is required to reverse decomposition of $LiBH_4$ [70]. Other systems involve the original $LiH/LiNH_2$ [66], more complex $2LiNH_2/MgH_2/LiBH_4$ composite, which decomposes in seven stages [71], or $MgH_2/Ca(BH_4)_2$ [53,65].

A very interesting system utilizing Kubas hydrogen binding: ($\eta^2$-$H_2$)–metal interaction, has been recently reported by Antonelli *and co-workers* [72,73]. This system, based on amorphous $MnH_x$ molecular sieve can be classified as falling in between metal hydrides and physisorption materials. It can be capable to reversibly store 10.5 wt% H, and almost 200 kg $m^{-3}$ at 120 bar and room temperature, with no loss of performance after more than 50 cycles. As this system remains close to thermo-neutral, it will certainly become an inspiration in further search for even better performing hydrogen storage materials. However, no detailed information has been given about the purity of $H_2$ gas used for absorption, and that coming from desorption, which raises questions about possibility of absorption of traces of water and/or $O_2$ by this compound. Similar findings have been reported for other related systems, like those based on chromium(III), where the capacity of *ca.* 5.1 wt% H was achieved [74].

## 2.2. Selection of binary and ternary intermetallics

### 2.2.1. *bcc* Ti-V alloys

Titanium alloy-based metal hydrides are an interesting class of hydrides because of their unique properties, their abundance and relatively low price. The volumetric hydrogen density of $TiH_2$ (150 kg $m^{-3}$) is more than two times that of liquid hydrogen. The formation enthalpy of $TiH_2$ is -123.4 kJ $mol^{-1}$ $H_2$ at 298 K [75], which means that significant heat is released during hydrogenation. This makes Ti and its alloys promising candidates for both on-board hydrogen storage and heat storage systems. The interaction of hydrogen with vanadium is rather complex in comparison to its reaction with titanium. The formation of various stable vanadium hydrides occurs below 200 °C. The formation enthalpy of *fcc* $VH_2$ is -40 kJ $mol^{-1}$ $H_2$ [76,77], hence these vanadium hydrides can be categorized as low-temperature hydrides.

Ti and V can be alloyed with various elements, e.g. transition metals Mn, Cr, Ni, Fe [78,79]. This produces either a solid solution alloy or an intermetallic alloy. The intermetallic alloy can have a *stoichiometric* ratio of *AB* and *AB$_2$*, where *A* is Ti or V, and *B* is usually a weak or non- hydride forming element. Many investigations have already been done on ternary and quaternary alloys [80-83]. Ti and V form continuous solid solution *bcc* alloys in the range 2.7 at.% to 80 at. % vanadium. *bcc* alloys can also be synthesized from Ti and V by the addition of beta stabilizer metals such as Cr and Nb. The addition of beta stabilizers has been known to reduce the thermal stability of the corresponding hydrides. The hydrogen capacity depends very much on the phase's constituent, the structure of the crystals, and the microstructure of the alloys. In this respect, Cr is favorable because Cr has a lower density than Ti and V. Addition of Cr to Ti-V based alloys is known to increase the equilibrium plateau pressure of the hydrides. This is related to the fact that Cr decreases the lattice parameter of the Ti-V-Cr *bcc* alloys since it has a lower metallic radius than Ti and V [81].

The *bcc* solid solution alloys are different from the intermetallic types in terms of ordering between the elements in the alloy. In *bcc* alloys, the alloying element is dissolved in the solvent such as Ti or V so that a disordered type alloy is formed [84]. On the other hand, the intermetallic alloys form ordered structures of two elements. *bcc* type alloys have not gained much attention apart from their inherent reasonably high hydrogen capacity. One reason for this is the slow activation process. However, Maeland *et al.* [79] showed that the *bcc* alloys absorb hydrogen fast if they are alloyed with Mo, Nb or Fe even without or only with slight activation process. The *bcc* type hydrides therefore became a promising alternative. Of the ternary system, several studies have been done on Ti-V-Cr based alloys in which the content of Cr and V were found to control the thermodynamics, kinetics, and hydrogen capacity [85-89]. Even though the crystal structure is *bcc* in most of those cases, an addition of another element may result in minor amounts of C14 Laves phase (hexagonal $MgZn_2$ structure) [90,91].

At low hydrogen concentrations, the binary Ti-V alloys form monohydrides where the structure is preserved and known as $\beta$-$Ti_{1-y}V_yH_x$ (where x ranges from 0 to ~1.4) with a lattice parameter expanding linearly with the H content [92]. The lattice parameter of $Ti_{1-y}V_y$ has been suggested to be linearly dependent on the amount of V at.%, due to differences in the atomic radius between Ti and V [80,93,94]. At high concentrations of hydrogen (H/M=1.45-2.00) hydrides of the $CaF_2$ type form, i.e. $\gamma$-$Ti_{1-y}V_yH_{1.50\text{-}}$. A *bct* structure can also be present when the alloy is rich in vanadium. The maximum hydrogen content of the *bcc* $Ti_{1-y}V_y$ alloy is 2.00 H/M or 3.5-4 wt.%. In the monohydride $\beta$-$Ti_{1-y}V_yH_x$, hydrogen occupies both *O* and *T* sites, and the population of the sites depends on the chemical composition of $Ti_{1-y}V_y$; the fraction of *T* sites occupied by hydrogen is higher in alloys with a high Ti fraction of [95,96].

In the intermetallic hydride class, $AB_2$ type alloys have gained a lot of attention because of their high hydrogen capacity compared to that of the $AB_5$ type alloys along with their easy activation [97-99]. According to Iba *et al.* [100], the composition of the alloy determines the phase-structure of the $AB_x$ alloy. Nevertheless, in the *A-B* system, 3 structural phases exist, namely *bcc*, C14 and C15 ($Cu_2Mg$ structure). The hydrogen capacity, thermodynamics, and kinetics of the bulk hydrides are affected by the fraction of each phase [101-103]. C14 and C15 typically have lower capacities than the *bcc* alloy. However, the C14 or C15 structures are considered beneficial to sustain the cycle ability and improve the activation properties of the alloys. In these regards, Iba and Akiba developed multiphase alloys which are better than individual Laves phases or *bcc* phase alone [100,104].

Hydrogen desorption from Ti-V hydrides was studied by thermal desorption spectroscopy [105]. The desorption spectra of hydrogen desorption are mainly composed of 3 peaks related to the stages of hydrogenation. The first step of dehydrogenation is low-temperature hydrogen desorption from the *fcc* $\gamma$-hydride. The hydrogen desorption then starts to evolve at about 300–400 °C depending on the amount of vanadium. This second mode of hydrogen desorption has the fastest hydrogen release rate. The last stage is the hydrogen desorption from *bcc* phase hydride.

The evolution of the phase-structural composition of the γ-$Ti_{0.8}V_{0.2}H_2$ (3.96 wt. % H) dihydride as a function of its thermal decomposition during a non-isothermal dehydrogenation was studied by *in situ* synchrotron radiation - powder X-ray diffraction (SR-XRD) and is shown in Fig. 2a. It can be seen as three-dimensional plot of the phase transformations of hydrides during dehydrogenation. The sample was heated from 25 °C to 800 °C with a heating rate of 2 °C min$^{-1}$ under vacuum. The diffraction patterns collected above 320 °C were fitted with a two-phase structural model, i.e. γ–*fcc* and δ- *bct* type hydrides. The reason for choosing this model was the observation of asymmetric profiles of the peaks corresponding to the (200), (220), and (311) of the γ- *fcc*. This is in line with a tetragonal distortion of the *fcc* lattice to a *bct* lattice. A proof that the sample was composed of the γ+δ hydride mixtures can be seen by observing the progression of the overlapped peaks γ–(111) and δ–(101) at temperatures from 450 °C to 500 °C, shown as an inset in Fig. 2a. Changes in phase abundances were calculated by the Rietveld method and can be seen in Fig. 2b. Phase transformations during the non-isothermal heating proceed according to the following pathways; γ–(hydride)→δ–(hydride)→β (hydride) →*bcc* alloy. The addition of Cr to these *bcc* Ti-V alloys changes the step of dehydrogenation to a multi-step process [106].

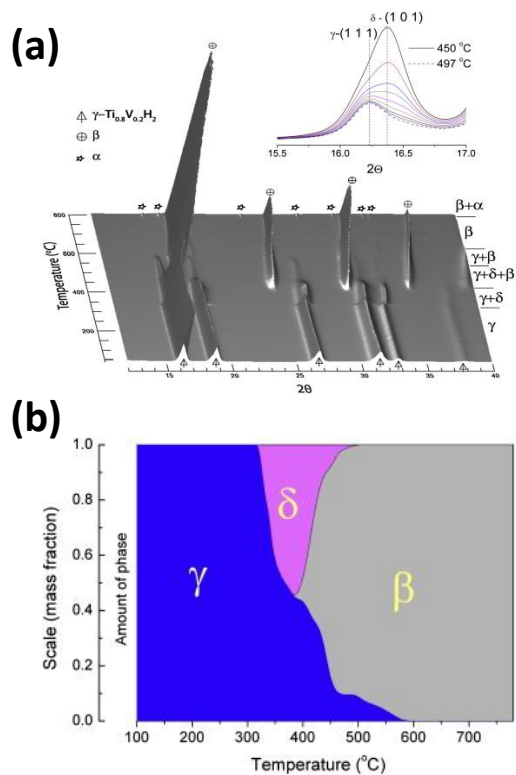

**Figure 2.**

(a) Phase evolution of γ-$Ti_{0.8}V_{0.2}H_2$ during an increase in temperature from 30 to 800 °C in vacuum studied by *in-situ* SR-XRD. The figure inset shows overlapping γ-(111) and δ-(101) peaks. Small fraction (~0.5 wt.%) of stable γ hydride was observed up to 580 °C. (b) Evolution of phase composition calculated from powder diffraction data of $Ti_{0.8}V_{0.2}H_x$ by Rietveld refinement method with Rwp = 3.76–10.28 % and Rp = 2.61–7.19 %. δ-hydride formed from γ-hydride at 320 °C and existed up to about 500 °C. The β hydride appeared at about 390 °C [105].

In principle, there are two routes for nano-structuring alloy for hydrogen storage, i.e. bottom-up approach with liquid-based synthesis method and bulk approach using ball milling, rapid solidification or severe deformation such as high-pressure torsion (HPT) method. There have been some studies on the effect of rapid solidification on *bcc* type alloys. Yu *et al.* [107] found that the rapid solidification of Ti-V-Mn-Cr increased the maximum hydrogen capacity. This was caused by a suppression of the C14 phase fraction in the alloy microstructure [107,108]. However, rapid solidification degraded the activation process, i.e. the first hydrogenation was slower as compared to the as-cast alloy, possibly due to oxide formation on the surface of the rapidly solidified samples. However, in another investigation, it was suggested that the existence of the C14 phase is helpful for the activation process because the C14 Laves phase easily cracks during cycling and provides paths for hydrogen penetration [91].

For *bcc* Ti-V, rapid solidification has improved the kinetics when a nanograin microstructure is formed [109,110]. The distribution of the alloying element is also affected by the nanostructured processing and enhance the hydrogen desorption properties. As can be seen in Fig. 3, the HPT processing results in an inhomogeneous distribution of Ti-V elements, which enhance the hydrogenation properties. Rapid solidification has been used to synthesize Laves phases in Zr based alloys [111,112]. It was observed that rapid solidification introduced a change in the phase fraction of C14 and C15 in the final microstructure. It was suggested that the formation of Laves phase microstructures do not only depend on the chemical composition alone, but also on the solidification rate [113]. Increasing the cooling rate increases the amount of C14 phase, but it leads to decreasing capacity. In addition, the activation process became much more difficult. One of the advantages was the prolonged cycle life of the alloys, i.e. up to 500 cycles without significant loss in capacity [111].

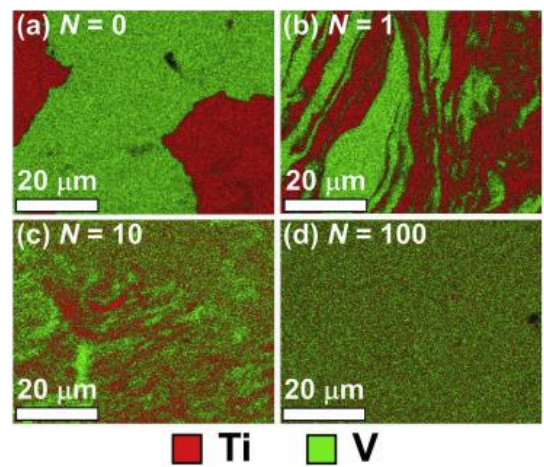

**Figure 3.**

Distribution of alloying elements after bulk nano-structuring in Ti-V alloys processed with high-pressure torsion (HPT) method. (a) to (d) the images indicate increased severity of deformation [114].

In general, the properties of metal hydrides can be altered upon hydrogenation-dehydrogenation cycling. This manifests itself by an alteration in the reversible capacity or the kinetic rate constant of the pressure-composition isotherm (PCI) profile. This change can be positive or negative for a specific application, however, in most cases the change is negative. The negative change or degradation can be categorized into two types:

- Intrinsic. The performance of the metal hydrides is decreased upon cycling due to the formation of more stable hydrides or compounds. This is the so-called disproportionation reaction.

- Extrinsic. The performance of the metal hydrides is decreased because of impurities in the hydrogen gas. These impurities are gases such as $CO_2$ that are reactive towards the hydride alloys.

For the case of Ti-V based alloys, cycling leads to pulverization and disproportionation of the alloys, which both decrease the reversible capacity. The *bcc* $Ti_{0.32}Cr_{0.43}V_{0.25}$ alloy has been shown to get lower particle size upon cycling, decreasing from 240 to 60 µm after 1000 cycles, and the hydrogen capacity was observed to decrease as well [115]. A rather worse case was observed in a *bcc* Ti-V-Cr-Mn alloy where the hydrogen reversible capacity was observed to decrease from 3.4 wt.% H in the $1^{st}$ cycle to only 2 wt.% after 200 cycles. Bulk analysis using XRD and surface studies using XPS revealed that the lattice volume of the *bcc* crystal was reduced and that the surface was enriched with oxygen as a consequence of the cycling [116,117]. This was suggested to cause the observed capacity decrease. A similar observation has been reported for V based alloys where pulverization has been suggested as the main reason for decreased capacity upon cycling [118]. In Ti-based alloys, the disproportionation is related to the formation of stable Ti hydrides, which have a higher thermal stability than the alloy. Observations of cycled samples using transmission electron microscopy (TEM) have clearly indicated that disproportionation occurs in $TiMn_2$ [119,120]. Improvements in the production process have contributed to increase the alloy performance upon cycling of Ti-V-Cr alloys [87], and composition homogeneity was suggested to be the reason for this increased cycling performance.

### 2.2.2. Ti-Fe based systems

Ti-Fe is another binary metal-based systems that has been extensively studied for hydrogen storage thanks to its tunable hydrogenation properties that can be tailored by elemental substitution, for its low cost, easy production and recycling [121]. The main investigated compound in the Ti-Fe system is the stoichiometric binary TiFe intermetallic. The first report on TiFe for hydrogen storage was published by Reilly *et al.* in 1974 [122], reporting a maximum gravimetric hydrogen capacity of 1.86 wt.% with full reversibility close to normal conditions of pressure and temperature. This feature makes TiFe suitable for stationary storage of hydrogen in mild conditions. TiFe can be produced by high temperature melting techniques, mechanical alloying, sputtering, or powder sintering.

The Ti-Fe binary phase diagram shows two intermetallic phases, TiFe and TiFe$_2$, which can be defined as *AB* and *AB*$_2$ compounds [123]. TiFe exists in a small range of composition, which extends from 49.7 to 52.5 at% Ti at 1085 °C. Both elemental Ti and Fe can form *bcc* solid solutions above 600 °C with reciprocal solubility up to approx. 10 at% of Ti in Fe at 1300 °C, and up to 20 at% of Fe in Ti at 1100 °C [124,125].

As most of Ti-based *AB* alloys, TiFe crystallizes in a CsCl-type cubic structure (S.G. *Pm-3m*). The enthalpy of formation of this cubic phase is –22.5 kJ mol$^{-1}$ [123]. The lattice parameter can vary because of composition fluctuation related to homogeneity domain of the TiFe phase that may arise from different synthesis methods [126]. Upon hydrogenation, TiFe forms two hydrides adopting orthorhombic structures: the monohydride TiFeH (β phase), and the dihydride TiFeH$_2$ (γ phase). Deuterated TiFe was investigated by neutron diffraction to define the hydrides structure. Consequently, the PCI curves under deuterium of TiFe present two distinguished plateau pressures that correspond to the β (1$^{st}$ plateau, Fig. 4) and γ-phase (2$^{nd}$ plateau, Fig. 4) [127]. Usually the two plateaus are better defined upon desorption, located at *ca.* 0.3 and 0.9 MPa at room temperature for the first and second plateau, respectively [128]. For absorption, the formation of a solid solution of H in *bcc*-TiFe is firstly observed (α phase). The solid solution α has a maximum H solubility at *H/M* = 0.04 (corresponding to TiFeH$_{0.08}$) [129]. Then, through the first plateau, the β phase forms (*P222$_1$*) [130], even if a large discussion concerning the TiFe hydride crystal structures is still present in the literature, together with some controversial results such as the orthorhombic structure (*C222*) reported in Ref. [131]. The dihydride, (γ phase) has an orthorhombic structure (*Cmmm*) [132]. The enthalpies of absorption Δ$H_{abs}$ have been reported equal to -17.5 kJ mol$^{-1}$$_H$ and -12.5 kJ mol$^{-1}$$_H$ for the α-β and β-γ transitions, respectively [126].

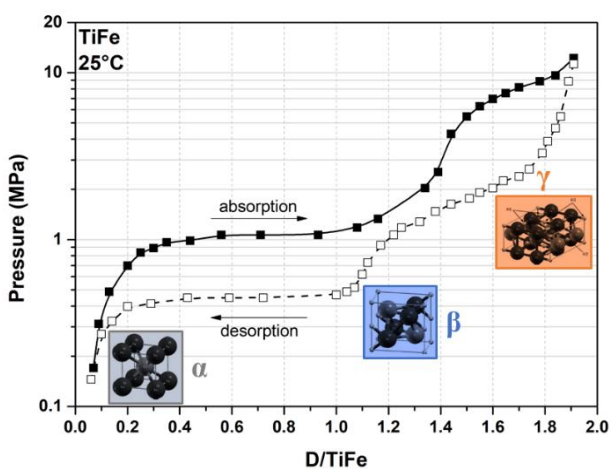

**Figure 4.**

TiFe PCI curves at 25 °C upon deuterium absorption (closed points) and desorption (open points), and respective crystal structure of the α, β, and γ phase. Solid line in absorption and dashed line in desorption are a guide for the eyes. Adapted from Ref. [127,131].

The cycling stability of the monohydride is good, while the dihydride has a sloppy plateau that shifts to higher pressures with increasing cycling. The activation of TiFe towards the first hydrogenation is usually challenging, since high temperature (above 400 °C) and long reaction time (days) are necessary for hydrogen atoms to overcome the native oxide layer at its surface. Moreover, TiFe has limited resistance to gas impurities. In addition, heat management and changes in volume (being approx. 18%) during absorption and desorption cycling must be considered [133]. In conclusion, TiFe is a low cost intermetallic with reversible hydrogen uptake near ambient conditions of pressure and temperature. However, the compound suffers from activation issues, sensitivity to gas impurities and cycling instabilities of the second plateau. Strategies to overcome these issues will be discussed in detail hereafter.

Frequently, as-synthetized and/or annealed TiFe does not react readily with hydrogen, neither at room temperature, nor under high hydrogen pressure (*i.e.* 10 MPa) [134]. The harsh activation of the stoichiometric single-phased TiFe alloy is attributed to the presence of native oxide passivating surface layers, since the intermetallic compound is sensible to contaminants such as water or oxygen. Air exposure creates a passivating surface layer, which prevents the hydrogenation. Precautions, such as using high hydrogen purity, must be taken for using TiFe as hydrogen storage medium. However, chemical substitutions and minor amounts of secondary phases at the TiFe grain boundaries are highly beneficial for the activation process, thus three main approaches can help the activation process in general.

The first one consists in cracking the metal-oxide layer by a high temperature (up to 400 °C) heat treatment in vacuum, in combination with applying high pressures of hydrogen (above 6 MPa), to facilitate as well hydrogen diffusion through the oxide layer and into the metal [122]. Furthermore, mechanical milling or grinding processes [134], or performing the absorption and desorption cycling under hydrogen enhance easy activation. By alternating cycles of absorption and desorption, the metal lattice is expanded during absorption and contracted during dehydrogenation. This results in pulverization of the metal particles, creating fresh metallic surfaces as well as a reduction in particle sizes. In this way, there is also the advantage to create a fine microstructure, which contributes in enhancing the kinetic of hydrogen absorption and desorption. Mechanical milling is also of interest as activation process. Indeed, milling can break the oxide layer, create fresh surfaces and refine the alloy microstructure [135]. However, extensive milling should be avoided to mitigate TiFe amorphization since it reduces its reversible storage capacity [136].

The second approach is to modify the chemical composition of the surface layer to precipitate either hydride-forming compounds or H-permeable phases allowing hydrogen penetration. As a matter of fact, if a secondary active phase such as Ti-rich precipitates are present at the alloy surface, its volume expansion due to the formation of Ti-rich hydrides (e.g. $TiH_2$) will favour cracking of the TiFe particles and expose fresh surfaces to accelerate hydrogen sorption. Pure Ti is expected to face the same oxidation issues, as $TiO_2$ also forms a passivation layer [137]. However, in the Ti-Fe system, the formation of Ti-rich precipitates containing iron ($Ti_{80}Fe_{20}$) instead of pure Ti, may help to activate the material [138]. Some oxide phases have been also reported to be reactive to hydrogen, allowing hydrogen diffusion towards TiFe and facilitating activation [139].

Finally, the third method to improve activation consists of chemical tailoring of composition. By modifying the binary composition of TiFe through suitable chemical substitutions either at the Ti or Fe sublattices, or for both elements, pseudo-binary intermetallic compounds more resistant to poisoning effect can be obtained requiring less laborious activation process [140,141]. It is important to note that, once activated; TiFe remains highly sensitive to atmospheric contaminants. Thus, for filling TiFe materials in hydrogen tanks, handling of activated TiFe powders under inert atmosphere (in purified glove box) should be considered. For large-scale applications, this approach is unpractical and therefore the activation process should be performed directly inside the tank. This implies that the operating conditions of the tank should be adapted to the activation conditions, with higher values of $H_2$ pressure and operative temperature than for the usual working conditions.

Addition of substitutional elements or secondary phases into TiFe compound allows the modification of its thermodynamic and chemical properties. Influence of metallic substitution on sorption properties of TiFe-based systems have been studied extensively and many different elements have been added to TiFe so far. By chemical substitutions, absorption and desorption pressures as well as the lengths of the plateaus are modified. Thus, chemical modifications allow tailoring performances and working conditions towards the final targets of the application. When the element addition results in the precipitation of some secondary phases, the reversible capacity is expected to decrease, especially when the secondary phases are not reversible or reactive to hydrogen at the operating conditions.

Tailoring TiFe means partial substitution of Fe at site 1*a* (0,0,0), or of Ti at site 1*b* (½,½,½) in the CsCl-type structure. According to Pearson's crystal database [142], 52 different substituted structures are reported, mostly compounds including one or two substitutions, with variations of the cell parameter in the range of 2.861 Å < *a* < 3.185 Å. The cell parameter of the pure equi-atomic Ti:Fe ratio is 2.976 Å. Introducing a substituent cause either the contraction or expansion of cell parameter and volume, depending on the size of the substituting element. For monosubstituted compounds, Fe can be replaced at site 1*a* in a large extent by Al, Co, Cu, Mn, Ni, Pd, whereas Ti can be substituted at site 1*b* by Al and Ru in a smaller compositional range. As for bi-substituted compounds, both 1*a* and 1*b* sites can be occupied by Al, Ga, Mn, Ru, and V. Monosubstituted compounds forming hydrides include α solid solution (Ru), β phase monohydride (Ni) and γ phase dihydride (Mn) [143-147].

It is worth noting that mixed site occupancies occur within the binary TiFe phase along the homogeneity domain, particularly for the Ti-rich side which extends up to 52.5 at% Ti. It was shown by accurate measurements of the density and the lattice parameter that the excess of titanium may substitute iron up to the homogeneity domain limit $Ti_{1b}(Fe_{1-x}Ti_x)_{1a}$ with $x = 0.05$ [148].

As mentioned before, hydrogenation of TiFe by a solid-gas reaction can only be achieved after severe activation due to its high sensitivity to surface poisoning by gaseous impurities [149,150]. However, partial substitutions of Fe by Mn and Ni are effective to improve the hydrogenation reactivity of the alloy, and allows adapting the plateau pressures for hydrogen storage applications [84]. Furthermore, the hydrogenation of titanium-rich alloys $Ti_{1+x}Fe$ ($0 < x < 0.05$), which contain β-Ti as a secondary phase, has been found to occur without any activation [151-154]. A previous study has shown that in certain cases where the Ti concentration is greater than 50 at%, the maximum hydrogen capacity exceeds that of the TiFe hydride [155]. This could be related to the presence of secondary phases, such as β-Ti precipitate as introduced before. In fact, $TiH_2$ has a capacity of 4 wt%, however it is not reversible under mild pressure and temperature conditions and, therefore, the reversible capacity of Ti-rich compounds above $x = 0.05$ is expected to be lower than for TiFe.

Mn substitution of Fe in TiFe has been reported to promote hydrogen uptake without activation treatment, especially in Ti-rich compounds such as $TiFe_{0.90}$. Both Mn and Ti in excess partially replace iron at the $1a$ site, causing an enlargement of the cell parameter and decreasing the plateau pressures [148]. Challet *et al.* reported experiments on $TiFe_{0.70}Mn_{0.20}$, evidencing chemical homogeneity and the absence of slopped plateaus (thanks to thermal treatment at 1000 °C for one week). Activation of the alloy at room temperature and 1 MPa after 2 hours of incubation time, and a hydrogen capacity of 1.98 wt% at room temperature and 2 MPa were observed. Two distinct plateaus were reported in the PCI curves. Furthermore, the initial capacity was retained over 20 cycles. Increasing the amount of Mn in the system produces a slight increase of the total capacity, better kinetics, overall decrease of both plateau pressures and a diminution in the difference of stability between mono and di-hydrides (first and second plateau desorption pressure) compared to Mn-free TiFe material [148,156].

By increasing the Fe/Mn ratio in $TiFe_{0.70+x}Mn_{0.20-x}$, the cell volume shrinks and, consequently, the plateau pressures rise. Thus, the plateau pressures at 65 °C of the materials investigated by Challet *et al.* are close to the pressure range 0.4-2 MPa. However, either the first plateau in desorption ($TiFe_{0.80}Mn_{0.10}$) or the one of the second hydride in absorption ($TiFe_{0.85}Mn_{0.05}$) are still out of this pressure domain. In addition, the pressure difference between the first and the second hydride increases with temperature, making it difficult to adjust both plateau pressures in a narrow pressure range. From the kinetic point of view, the alloys exhibit very fast reaction rate since 90% of the total capacity is absorbed in less than 3 minutes at room temperature and 2 MPa.

V substitution of Ti in TiFe-systems has also evidenced an improved and easy activation process at room temperature and 2.5 MPa of hydrogen, with an increase of the capacity as a function of V content, and no clear plateau pressure difference between mono and di-hydride. However, PCI curves show sloppy plateaus [157].

Chromium and aluminium can also be considered in monosubstituted compounds. A mechanical-substituted alloy, with 4 at% Cr and 5 at% Al respectively, shown a reversible capacity of 0.7 wt% at room temperature and between 0.1-10 MPa. In these systems, Al shows a higher plateau pressure compared to Cr-substituted material, and both remarkably reduce hysteresis with respect to pure TiFe, improving also kinetics and activation [158].

Finally, the partial substitution of Fe with Ni, Mn, V, Cr, Al, Co can decrease the first plateau pressure and facilitate the activation [11,84,148,156-159]. The plateau pressure decreases considerably passing from non-substituted alloy to $TiFe_{0.9}Ni_{0.1}$. The plateau remains as flat as for TiFe, except for $TiFe_{0.9}Al_{0.1}$. In this case, the substitution creates a stiff slope which is attributed to the distortion of octahedral sites due to the large difference of metallic radii between Fe and Al ($r_{Fe}$ = 1.24 Å, $r_{Al}$ = 1.43 Å) [160]. It is hypothesized that it exists a size distribution of octahedral interstitial sites with different hydrogen occupancies. Such inhomogeneity would enhance the slope. Thus, to avoid this issue, best substituting *B*-type elements would be those with metallic radius comparable to Fe. This kind of local effects have been suggested in the past for $AB_2$ compounds as well [161,162]. However, this effect seems not to affect the slope of the plateau in case of the Ti-rich TiFe compounds, where no sloping plateau is observed and $r_{Ti}$ = 1.47 Å, a radius even higher than that of Al.

The remarkable decrease of the plateau pressure with Ni and Co substitutions is linked to the stronger affinity towards hydrogen of TiCo and TiNi with respect to TiFe, which implies a higher enthalpy of formation; for $TiNiH_{0.9}$ equal to -30 kJ mol$^{-1}$H$_2$ and for $TiCoH_{0.9}$ equal to -27 kJ mol$^{-1}$ H$_2$ [163], resulting in more stable monohydrides. Substitutions can also act on decreasing hysteresis phenomenon among cycles; particularly effective are Ni and Mn elements. However, the hysteresis loop is affected by the initial temperature and pressure [164].

In conclusion, the variety of ternary substitutions in TiFe intermetallic compounds demonstrates how flexible this material is and how much it can be tuned towards the final application. In most of the cases, substitution is beneficial and improves properties such as hydrogen capacity, low hysteresis, kinetics and easy activation. Some elements have drastic effects and may also modify the pressure step between the first and second plateau. The plateau pressure can be finely tuned as a function of the quantity of substituent introduced. For this reason, thanks to its low cost and easy processing, TiFe substituted systems are promising for large-scale stationary application, and can be easily tuned and coupled with electrolyzer and fuel cells incrementing the spread-out of hydrogen as efficient energy carrier and supporting the development and integration of renewable energies into smart grids.

## 2.3. High entropy systems

Multi-principal element alloys (MPEAs), also known as high-entropy alloys (HEAs), are a relatively new class of materials constituted by several elements in near equimolar concentrations initially reported independently by two groups in 2004 [165,166]. In these works, the authors studied two quinary systems, exploring the central region of the multi-component phase diagram as a new approach for alloy development. Interestingly, these alloys are disordered solid solution which can adopt simple crystalline structures (*bcc*, *fcc* and *hcp*). It has been suggested that the high configurational entropy is the cause of the stabilization of the single-phase solid solutions. Behind the HEA term, it is now agreed that the motivation is to find single-phased solid solutions by controlling the configurational entropy [167]. On the contrary, the MPEA term is employed to remind the vastness of composition space of the central region of phase diagram regardless the entropy contribution or the number and types of phases present. In this respect, the MPEA includes a broader classification of alloys then HEA, but the latter term has acquired a more widespread recognition. In addition, it appears that there are two commonly accepted definitions for HEA, one based on the magnitude of entropy, and another based on the concentration of principal elements. The entropy-based definition imposes the entropic term of a disordered solid solution to be superior to $1.61R$ ($R$ is the universal gas constant), whereas the definition based on composition emphasizes that the concentration of elemental constituents should be in the range 5-35 at.%.

This new strategy of alloy design is very exciting due to the vast number of possible configurations and has become a playground of many research activities mainly focused on the structural and mechanical properties. In addition, other interesting functional properties, such as thermoelectric effect, photovoltaic conversion, piezoelectricity have been reported. Hydrogen storage in HEA is particularly interesting since these alloys are composed of elements with different atomic radii and thus a high lattice strain is developed possibly providing large interstitial sites for hydrogen occupation. However, the hydrogen sorption properties of HEA are scarcely reported in literature. The mainstream of reports focus on *bcc* alloys [168-176] and hexagonal C14 Laves phases [177-180]. The present review highlights hydrogen storage properties of refractory *bcc* alloys since this class of materials in their conventional form (see § 2.2) is well known for hydrogen storage for many years [181].

In 2014, Kunce *et al.* [168] investigated the hydrogen sorption properties of TiZrNbMoV synthesized by laser engineered net shaping. The single-phased *bcc* alloy showed a low hydrogen capacity of 0.6 wt.% at room temperature. Later, Sahlberg *et al.* [169] reported the hydrogen sorption properties of a refractory HEA TiVZrNbHf showing an outstanding capacity. Interestingly, this alloy has a hydrogen uptake of 2.5 H/*M*, which is greater than the capacity of individual elements (2 H/*M*). Moreover, the hydrogenation of this alloy is a reversible single step reaction (*bcc* ↔ *bct*). Such high hydrogen content has never been observed in individual hydrides based on transition metals and this implies that hydrogen occupies not only tetrahedral but also the octahedral interstices in the *bct* (pseudo-*fcc*) lattice [170].

Zepon *et al.* [171] synthesized MgZrTiFe$_{0.5}$Co$_{0.5}$Ni$_{0.5}$ using high-energy ball milling, a technique suitable for the preparation of alloys containing high vapor pressure metals, such as Mg. In their work, the alloy synthesized under Ar atmosphere crystallizes in a *bcc* structure and has a maximum hydrogen absorption of 1.2 wt.% (0.7 H/*M*) at relatively high temperature. A reversible transformation from (pristine) *bcc* ↔ (hydride) *fcc* structure upon hydrogenation was proven by *in-situ* synchrotron diffraction. Later, Zlotea *et al.* [172] reported the hydrogen storage properties of the single-phase *bcc* TiZrNbHfTa alloy synthesized by arc melting. The pressure-composition isotherm shows two plateaus, one at low equilibrium pressure reaching a capacity of 1 H/*M*, followed by a second plateau at 23 bar with a maximum capacity of 2 H/*M* (1.7 wt.%). SR-XRD experiments prove that this alloy has a two-step reversible transformation from (pristine) *bcc* ↔ (monohydride) *bct* ↔ (dihydride) *fcc* phase. This behavior contrasts with the previously reported refractory HEA with a single hydrogenation step but in agreement with the conventional *bcc* alloys [181]. Recently, Montero *et al.* reported a comparative study on the hydrogen sorption properties for Ti-V-Zr-Nb alloys synthesized by three different methods: arc melting, high-energy ball milling in Ar atmosphere and reactive ball milling under high $H_2$ atmosphere [175]. The purpose of this study was to determine the most suitable technique for the refractory HEA preparation. The first two methods yielded a single-phase *bcc* alloy, while the synthesis by reactive ball milling directly forms a hydride phase with pseudo-*fcc* lattice. In terms of capacity, the three materials showed similar hydrogen uptake, 1.7-1.8 H/*M* (2.6-2.7 wt.%), however, the ball-milled sample showed the slowest absorption kinetics and the highest temperature of desorption. Additionally, the hydrogen absorption/desorption cycling was reported over 20 cycles, such a property is still lacking in the literature. After an initial fading of the capacity during the first cycles, this alloy was able to reversibly store a stable capacity of 1.3 H/*M* (2.0 wt.%) up to 20 cycles. As an extension of this work, Montero *et al.* have very recently shown that the addition of only 10 at.% of Ta into the quaternary Ti-V-Zr-Nb alloy reported earlier improves the hydrogen sorption performances [182]. The absorption/desorption in the quinary $Ti_{0.30}V_{0.25}Zr_{0.10}Nb_{0.25}Ta_{0.10}$ alloy is reversible and occurs within one step (*bcc* alloy ↔ *fcc* dihydride). The capacity is only slightly fading during the first 10 cycles and then stabilizes at around 1.7 H/*M* (2.2 wt.%) for the next 10 cycles.

In a different approach towards the understanding of the hydrogen sorption properties of HEA, Nygård *et al.* studied a series of *bcc* alloys with a scaling degree of lattice distortion [173]. Several alloys in the form $TiVZr_{(1-z)}NbTa_{(z)}$ and $TiVZr_{(1+z)}Nb$ were synthesized in which the amount of Zr relative to the alloy ([Zr]/[*M*]) was controlled in order to tune the lattice distortion. Most of the alloys have a maximum uptake of 1.8-2.0 H/*M* but no clear relationship is observed between the hydrogen capacity and the lattice distortion of the HEA nor the lattice parameter *a*. The stability of the hydride increased as [Zr]/[*M*] increases. Interestingly, alloys with [Zr]/[*M*] < 12.5 at.% have recovered their initial *bcc* structure after desorption, while those with [Zr]/[*M*] > 12.5 at.% have suffered phase separation after one cycle of hydrogen absorption/desorption. The same group reported a second study of various ternary, quaternary, and quinary refractory alloys with different valence electron concentration (VEC) [174]. In this work, they found a linear relationship between the hydride stability and the VEC. Moreover, the electron concentration is already known to affect the hydrogen storage capacity of classical *bcc* alloys; the increase of this parameter above a certain limit induces a drastic decrease of the capacity [183]. Hydrogen cycling experiments have proven a drop of capacity to zero absorption for most of the alloys after few cycles, with one exception TiVCrNb, which losses capacity from 3 to 2 wt.% after the first cycle and then stabilizes to this value for further cycling. Very recently, Shen *et al.* have studied the single-phased *bcc* TiZrHfMoNb alloys with varied Mo content [176]. These alloys absorb hydrogen reversibly forming hydrides with *fcc* lattice. The thermal stability of the hydrides decreases with increasing Mo concentration, mainly attributed to smaller cell volume by Mo addition.

The majority of these refractory *bcc* alloys rapidly absorb hydrogen within a single step reversible reaction *bcc* ↔ (pseudo) *fcc* with an equilibrium pressure below 1 bar (Fig. 5, left). However, the hydrides are very stable and high temperature is needed to desorb the hydrogen, typically 300 °C, under vacuum. Therefore, the main challenge in the future is to destabilize HEA hydrides by increasing the equilibrium pressure above 1 bar at room temperature in order to have access to the whole length of the plateau, *i.e.* a useable capacity of 2 H/*M* available by playing with the pressure with the dotted box in Fig. 5 (right).

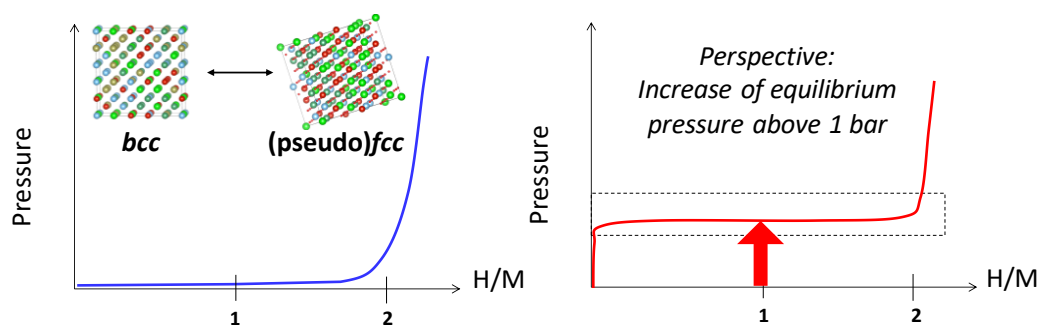

**Figure 5.**

Most common Pressure-Composition-Isotherms (PCIs) for hydrogen absorption in *bcc* HEAs with equilibrium pressure below 1 bar (left) and proposed PCIs with equilibrium pressure above 1 bar to have a useable capacity of 2 H/*M* (right).

Finally, the study of hydrogen absorption/desorption properties of this new class of alloys is at an early stage but the results in the literature are promising. Substantial research efforts are still needed in order to rationalize the behavior of these alloys towards hydrogen due to the extensive number of possible elemental combinations in terms of chemical composition, elemental concentration, VEC and lattice distortion.

## 2.4. Borohydrides of alkali and alkaline earth metals

Sodium borohydride, produced on a scale of a few thousand metric tons annually, remains the most important industrially useful complex metal hydride [184], while also a few other borohydrides are commercially available. The list originally included only the M(BH$_4$)$_n$ salts of alkali metals, M = Li – Cs; n=1, and a few years ago it has been expanded by their analogues containing Mg and Ca, n=2. Among these compounds mostly NaBH$_4$ and LiBH$_4$ became the basis for the impressive extension of the number of known solvent-free borohydrides to *ca.* 150, counting only those of identified crystal structures [185,186]. This remarkable expansion of chemistry of borohydrides observed during the last two decades has been mainly inspired by their potential for hydrogen storage [186-188]. In this section, we will review some of the recent developments concerning the borohydrides of metals of the first two groups of the periodic table.

The borohydrides of alkali metals, MBH$_4$, clearly reveal an ionic character – they all crystalize in disordered CsClO$_4$-type structure of tetrahedral coordination of metal cation by the borohydride groups (NaCl structural type, if the heavy atoms are only considered), although for LiBH$_4$ this type of crystal structure is observed only above 10 GPa [189]. At ambient conditions, the latter compound forms an orthorhombic crystal (*o*-LiBH$_4$) with tetrahedral arrangement of BH$_4^-$ around Li$^+$, while at *ca.* 110 °C a phase transition to a hexagonal, wurtzite-like, polymorph occurs (*h*-LiBH$_4$)[190,191]. This hexagonal phase reveals increased solid-state Li$^+$ conductivity, and could be stabilized at room temperature by halide substitution. Thus, the solid solutions like *h*-Li(BH$_4$)$_{0.7}$Br$_{0.2}$Cl$_{0.1}$ or Li(BH$_4$)$_{0.66}$I$_{0.33}$ show room-temperature conductivity in the order of 10$^{-5}$ S cm$^{-1}$ [192,193]. More details about solid-state electrolytes are given in Section 4.

The solid-state structures of alkaline earth borohydrides, M(BH$_4$)$_2$, are influenced by partial covalent interactions observed in these compounds. These manifest most clearly for the lighter elements: The Be(BH$_4$)$_2$ crystal is built of helical chains propagating along the *c* direction of the unit cell [194] while Mg(BH$_4$)$_2$ and Ca(BH$_4$)$_2$ reveal very broad structural diversity, with several metastable polymorphs detected at ambient conditions some of them showing rather loose packing [185,186]. Among these, γ-Mg(BH$_4$)$_2$ with *ca.* 33% empty volume is the most impressive example of borohydride-based porous material, as it reversibly absorbs H$_2$, N$_2$ or even CH$_2$Cl$_2$ [195].

The most convenient method to prepare alkali and alkaline earth metal borohydrides are various solution-based reactions with the exclusion of small samples for research purposes. However, a few interesting, mechanochemical approaches have been reported. One of them is the preparation of Ca(BH$_4$)$_2$ from synthetic colemanite – one of the important mineral sources of boron [196]:

$$Ca_2B_6O_{11} + 12CaH_2 \rightarrow 3Ca(BH_4)_2 + 11CaO \qquad (3)$$

Utilizing the Mg–Al-based waste in mechanochemically-induced reactions with the respective boranes in hydrogen atmosphere leads to an efficient synthesis of LiBH$_4$ or NaBH$_4$ [197], *e.g.*:

$$NaBO_2 + 2Mg + 2H_2 \rightarrow NaBH_4 + 2MgO \qquad (4)$$

These two reactions described in eq. (3) and (4) are certainly useful to obtain the desired product in a simplified way, straight from the natural sources of boron.

While NaBH$_4$ and LiBH$_4$ are available on a large scale via well-established methods [184], the heavier MBH$_4$, M = K – Cs, can be prepared in the ion metathesis reaction performed in cold methanol [198]:

$$\text{MOH} + \text{NaBH}_4 \rightarrow \text{MBH}_4\downarrow + \text{NaOH} \quad (5)$$

Although this method seems convenient even on a larger scale, it requires well-established protective measures. As methanol is a protic solvent, partial solvolysis of borohydrides is difficult to avoid and may even lead to the explosion of the reaction vessel (!) due to an abrupt pressure increase if no depressurization devices have been used.

The alkali metal borohydrides can be obtained from solvent-mediated metathetic reactions between the halide of corresponding divalent metal and lithium or sodium borohydride, followed by desolvation [199,200]. Contrastingly, toxic Be(BH$_4$)$_2$, due to its volatility, forms already during heating of such mixture to *ca.* 155 °C and can be purified by vacuum sublimation [194]. Another approach involves the reaction between a metal hydride and borane complexes such as (C$_2$H$_5$)$_3$N or (CH$_3$)$_2$S, eq. (6), which has been utilized for the preparation of M(BH$_4$)$_2$, M = Mg – Ba, and other borohydrides [199,201,202].

$$\text{MgH}_2 + 2(\text{C}_2\text{H}_5)_3\text{N}\cdot\text{BH}_3 \rightarrow \text{Mg(BH}_4)_2 + 2(\text{C}_2\text{H}_5)_3\text{N} \quad (6)$$

As it has been discussed for simple metal hydrides, thermal stability of these compounds can be rationalized on the basis of electron acceptor properties of the metal atom, as probed by the respective standard redox potential [31]. A similar correlation has been found between T$_{dec}$ of borohydrides, which raise with decreasing Pauling electronegativity of the metal atom [187,203]. Consequently, it can be expected that borohydrides of alkali and alkaline earth metals will be the most thermally stable. Indeed, besides the borohydride of relatively electropositive beryllium (T$_{dec}$ of *ca.* 123 °C), the borohydrides of other metals considered in this section decompose above 300 °C, which is very far from the temperature required for systems coupled with a PEM fuel cell [30]. At the same time, the reversibility of storage (*i.e.* possibility of hydrogen charging and re- charging) remains another important issue which does not show satisfactory parameters for the aforementioned borohydrides. Therefore, plenty of tuning options have been tested for these and related systems. These include forming of the reactive hydride composites discussed above, testing various catalytic approaches, *e.g.* nucleation by amorphous boron [204], nanoconfinement using broad range of scaffolds and particle sizes [205,206], or modification of chemical composition via preparation of mixed-cation or mixed-anion compounds [207,208].

During the last two decades, Mg(BH$_4$)$_2$ was mostly considered for its high hydrogen mass content and its low predicted decomposition temperatures. The mass composition of 14.94% H next to 45.02% Mg and 40.05% B, shows among the highest known hydrogen contents. However, the prediction that reversible hydrogen storage could be achieved at temperatures [209,210] suitable for proton-exchange membrane (PEM) fuel cells has not been fulfilled until today. That would have made this compound one of the most favorable materials for solid-state hydrogen storage. In practice, reversibility was achieved but never in respect to the full hydrogen content.

From a structural point of view Mg(BH$_4$)$_2$ is very interesting, and a number of different polymorphs have been reported. Some crystal structures that are of interest for the dynamics are shown in Fig. 6. Currently, there are seven known polymorphs, which is the largest variety of structures among the alkali and alkaline-earth borohydrides. Five of these structures have been identified (α-, β-, γ-, δ-, ζ-Mg(BH$_4$)$_2$) and two experimentally found but structurally unsolved (β'- and ε-Mg(BH$_4$)$_2$). Additionally, Mg(BH$_4$)$_2$ can form an X-ray amorphous material, which can be obtained by different methods. These methods are solvent-free synthesis [211], ball milling of the α- and γ-modifications [195] and a pressure collapse of γ-Mg(BH$_4$)$_2$ [212]. The latter method also resulted in a novel phase, δ-Mg(BH$_4$)$_2$ [195]. The δ-polymorph has received special attention as it has one of the highest volumetric hydrogen capacities of all complex metal hydrides with 147 kg m$^{-3}$. The other phases have volumetric hydrogen capacities between 82 - 117 kg m$^{-3}$ as well.

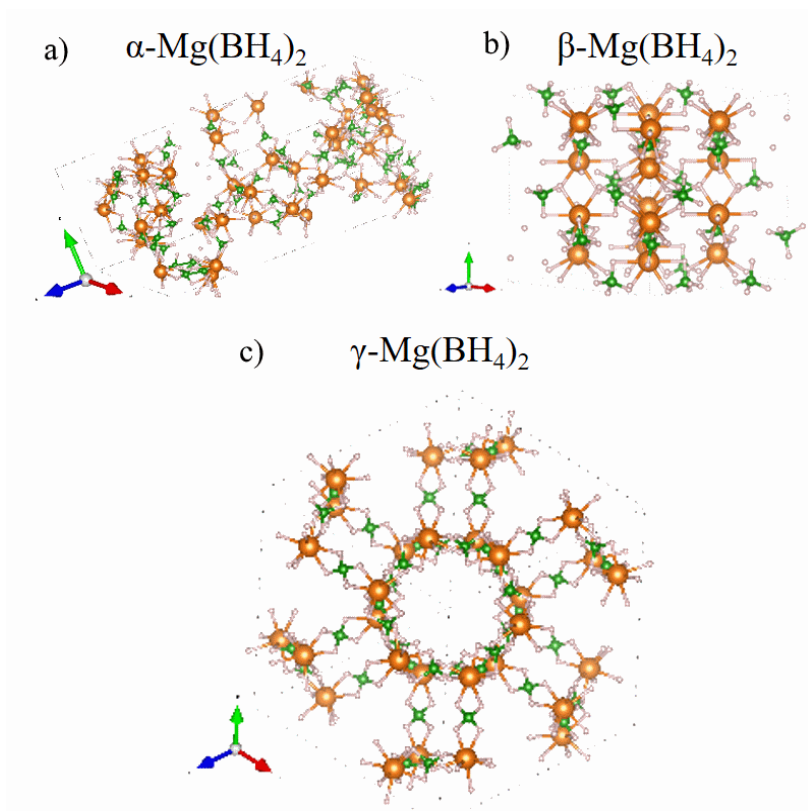

**Figure 6.**
The most important structures for this chapter are shown a) α- Mg(BH$_4$)$_2$ in space group (SG) P6$_1$22 [213], b) β-Mg(BH$_4$)$_2$ in SG Fddd [214], c) γ- Mg(BH$_4$)$_2$ in SG Id−3a [195,215].

From diffraction the highly symmetric γ-Mg(BH$_4$)$_2$ phase in space group Id–3a was determined by Filinchuk *et al.* [195]. One Mg atom is coordinated by the edges of four tetrahedral anions [BH$_4$]$^-$. It was reported that γ-Mg(BH$_4$)$_2$ has a three-dimensional mesh of interpenetrating channels. These have an inner and outer diameter of ~8 and ~12.3 Å, respectively. Therefore, this compound can almost be seen as a metal organic framework-like-structure with a porosity of ~33%, which is attractive to adsorb guest species [195]. Filinchuk *et al.* reported that 3 wt.% of hydrogen can be adsorbed at −193 °C and 105 bar of hydrogen. Moreover, nitrogen and dichloromethane have been reversibly adsorbed.

Heere *et al.* recently investigated the structure of the amorphous phase and the results are further discussed in Ref. [215]. The study used X-ray total scattering and pair distribution function (PDF) analysis, where even hydrogen bonds can be observed, due to the circumstance that H has an oxidation state of –1 and with that a notable electron density. The complete picture (or hydrogen positions) of the structure might not be given, nevertheless, the PDF analysis of the amorphous structure strongly resembles the one of γ-Mg(BH$_4$)$_2$. In detail, the PDF can be divided into the local structure up to 5.1 Å with the building blocks Mg – BH$_4$ – Mg. These results are in agreement with the findings by Filinchuk *et al.* whose spectroscopic results pointed to characteristic building blocks in the X-ray amorphous phase and their similarity to the internal structure of the γ- and δ-Mg(BH$_4$)$_2$ [195]. Above ~5.1 Å, a slight oscillation in the PDF is still recognizable, which Heere *et al.* connected to the interpenetrating channels, see Fig 6c, thus indicating that the fundamental building blocks still exhibit a certain degree of ordering related to the 3D net of interpenetrating channels, although less porous [195,216]. Above 12.3 Å, i.e. the outer diameter of these channels, the featureless PDF supports a fully randomized structure.

The thermal behavior prior to decomposition is characterized by DSC measurement. It was reported that there are endothermic and exothermic phase changes, which are all specific to the respective crystal phases. The first reported was the transition from α-Mg(BH$_4$)$_2$ to β-Mg(BH$_4$)$_2$ during thermal treatment at ~220 °C. β-Mg(BH$_4$)$_2$ is the high temperature modification which is metastable at room temperature and therefore not reversible. DSC of the γ-Mg(BH$_4$)$_2$ compound reveals two distinct phase transitions. The first one at ~150 °C is an endothermic event to ε-Mg(BH$_4$)$_2$, while the second event at 180 °C belongs to the transformation to β'-Mg(BH$_4$)$_2$ [215,216]. The amorphous phase synthesized via a ball milling reaction [217] of γ-Mg(BH$_4$)$_2$ shows an additional exothermic DSC event at 100 °C, where the material crystallizes to γ-Mg(BH$_4$)$_2$. The other two events are endothermic and related to the ones of the original γ-Mg(BH$_4$)$_2$, though lower in peak temperatures. At temperatures over 300 °C the decomposition of Mg(BH$_4$)$_2$ appears with four reported reaction steps [218]. At these temperatures, very stable compounds are formed, such as MgB$_{12}$H$_{12}$ and MgB$_2$, which can only be rehydrogenated at extreme conditions of 400 °C and 900 bar H$_2$ [219]. More suitable conditions for on-board hydrogen storage can be achieved via an intermediate step and the formation of Mg(B$_3$H$_8$)$_2$ at T ~200 °C. Hence a rehydrogenation to Mg(BH$_4$)$_2$ was achieved after 48 h at 250 °C and 120 bar H$_2$ [220]. The authors reported further improvement of the conditions and a rehydrogenation of a complex of $Mg(B_3H_8)_2 \cdot 2THF/MgH_2$ after 2 h at 200 °C and 50 bar $H_2$ [221], with the reaction products being discussed in Ref. [222]. As mentioned above, reversibility was achieved due to the intermediate product and the reaction pathway from $Mg(B_3H_8)_2$ to $Mg(BH_4)_2$, but never in respect to the full hydrogen content. The hydrogen capacity for this reaction is limited to 2.5 wt.% and the reaction conditions are still too high (~120 bar $H_2$, $T_{Absorption}$= 260 °C for 7-8 h, or 280 °C for 3 h) for on-board hydrogen storage [223].

## 2.5. Borohydrides of transition and rare-earth metals

The most emphasized aspects in the research devoted to borohydrides involve their synthesis, identification and physicochemical characterization, including crystal structures, spectroscopic measurements, magnetism, luminescence and ionic transport, as well as the processes related to hydrogen release. In this section we will briefly discuss some of these topics on a basis of various borohydride systems studied recently, mostly based on transition- and rare-earth metals, as well as some less common systems, like those containing ammonium cations.

### 2.5.1. Synthesis and basic structural identification

The rapid growth in the field of borohydrides, mentioned in the previous section, would not be possible without the development of a universal synthetic methodology facilitating fast screening of a broad range of compounds to assess their potential use for hydrogen storage. Consequently, the mechanochemical reactions performed in solid state *via* high-energy milling (using balls or discs) became the "golden standard" for synthesis of novel borohydrides [135,217,224,225]. This approach allowed for performing either ion metathesis, eq. (7), or salt addition, eq. (8), resulting in either single- and multiple-cation compounds, as exemplified below [226-235]:

$$YCl_3 + 3LiBH_4 \rightarrow Y(BH_4)_3 + 3LiCl \qquad (7)$$

$$Y(BH_4)_3 + KBH_4 \rightarrow K[Y(BH_4)_4] \qquad (8)$$

Using the mechanochemically-induced solid-state reactions even the compounds of fairly complicated stoichiometries can be obtained, like $Li_3MZn_5(BH_4)_{15}$, M = Mg, Mn, $Li_3K_3M_2(BH_4)_{12}$, M = La, Ce or $M_2LiY(BH_4)_6$, M = Rb, Cs [234,236-239]. This method is also an obvious choice for introducing doping or preparation of composites due to easily achieved very high degree of dispersion [53,65]. Identity of the prepared products may depend on the stoichiometry of the initial mixture of precursors, like for $NaZn_2(BH_4)_5$ and $NaZn(BH_4)_3$ obtained from the $ZnCl_2$-$LiBH_4$ mixtures of 1:2.5 and 1:3 ratio, respectively or for the borohydrides of the heavy alkali metals and yttrium [234,240]. Interestingly, the ratio of reagents can also determine which polymorph of the product is formed. This has been observed to some extent in the case of Cd(BH$_4$)$_2$ [241], and allowed for selection of the polymorph for the series of rare earth (*RE*) borohydrides, according to eq. (9) and (10), enabling to study of structure-related properties [242-244]:

$$RECl_3 + 3LiBH_4 \rightarrow \alpha\text{-}RE(BH_4)_3 + 3LiCl \qquad (9)$$

$$RECl_3 + 12LiBH_4 \rightarrow \beta\text{-}RE(BH_4)_3 + 3LiCl + 9LiBH_4 \qquad (10)$$

A systematic screening of the early transition metal borohydrides has recently been performed using a convenient mechanochemical approach [245,246]. The screening included the borohydride systems based on combinations of Ti, V, Cr, Mn, Fe with alkali metals, prepared according to eq. (11):

$$3MBH_4 + M'Cl_n + nLiBH_4 \rightarrow M_3M'(BH_4)_{3+n} + nLiCl \qquad (11)$$

where M = K – Cs, M' = Ti – Fe, *n* = 2 or 3. While no signs of a reaction were observed for vanadium, the expected borohydrides have been detected for other early transition metals and Fe, mainly in combination with Rb and Cs. For the borohydrides of lighter alkali metals either different products were formed, as in the case of K$_2$Mn(BH$_4$)$_4$, or no reactions occurred for other tested systems. It is worth mentioning that no borohydrides of trivalent metals formed, as in these systems either reaction were not observed (as for V(III)), or it occurs with subsequent decomposition of the products (as for Fe(III)), or with reduction to the respective divalent form (as for Ti(III)). A similar synthetic path (*i.e.* mechanochemical reactions performed at room temperature or <0 °C) has also been attempted for the monometallic borohydrides of M', however, it succeeded only for Mn(BH$_4$)$_2$ [247]. The identity of compounds prepared has been confirmed by the refinement of their crystal structures and are tabulated in Table 2. These borohydrides adopt Cs$_3$CoCl$_5$-type structure shared by M$_3$Mg(BH$_4$)$_5$ compounds, M = K (only above 94 °C) [248], Rb and Cs [249]. Contrastingly, Rb$_3$Cr(BH$_4$)$_5$ crystallizes in the structure related to room-temperature modification of K$_3$Mg(BH$_4$)$_5$ [248]. Thermal decomposition of these compounds occurs via exothermic processes within the range of temperatures as low as 50 – 120 °C, with emission of markedly pure hydrogen. For M' = Mn hydrogen contaminated with diborane is released at slightly higher temperatures.

Table 2. The parameters of the crystal unit cells of M$_3$M'(BH$_4$)$_5$ compounds prepared mechanochemically according to eq. (11). * - data measured at –43 °C due to limited thermal stability, the other diffraction data obtained at ca. 25 °C. The compounds adopt tetragonal crystal structures related to that of – and usually isostructural with – K$_3$Mg(BH$_4$)$_5$ [248]. Estimated standard deviations in parentheses.

| Formula | group | *a* [Å] | *c* [Å] | *V* [Å$^3$] | ICSD # |
|---|---|---|---|---|---|
| **Rb$_3$Ti(BH$_4$)$_5$** | *I*4/mcm | 9.214(3) | 16.130(5) | 1369.4(10) | 1990334 |
| **Cs$_3$Ti(BH$_4$)$_5$** | *I*4/mcm | 9.644(8) | 16.426(15) | 1528(2) | 1990346 |
| **Rb$_3$Cr(BH$_4$)$_5$** | *P*4$_2$/mbc | 9.221(3) | 16.275(6) | 1383.7(9) | 1990401 |
| **Rb$_3$Mn(BH$_4$)$_5$** | *I*4/mcm | 9.2963(19) | 16.101(3) | 1391.5(5) | 1990086 |
| **Cs₃Mn(BH₄)₅** | *I*4/mcm | 9.716(2) | 16.354(4) | 1543.8(7) | 1990328 |
| **Rb₃Fe(BH₄)₅*** | *I*4/mcm | 9.171(4) | 15.833(6) | 1331.7(13) | 1990394 |
| **Cs₃Fe(BH₄)₅*** | *I*4/mcm | 9.619(11) | 16.014(19) | 1482(3) | 1990400 |

Although convenient for diverse systems, less laborious than the solvent-mediated procedures and technically simple on a small scale, which strongly increased its popularity, the mechanochemical method has also some serious drawbacks. The most important is the purity of the borohydride product, which is severely limited by the "dead mass" of the halide by-product contributing to up to 50 wt.% of the post-reaction composite, and thus dramatically decreases the effective hydrogen content. Most borohydrides cannot be purified using typical solvents like ethers and amines or by physical methods, like difference in buoyancy, due to formation of very stable solvates and high dispersion, respectively [250-253]. Moreover, the side reactions connected with the formation of (BH$_4$/Cl) solid solutions may be observed, often influencing the process of thermal decomposition [231,234,254,255] and some of the borohydrides are still inaccessible *via* mechanochemically-induced solid state reactions [251].

These disadvantages stimulated the search for more general solvent-mediated synthetic approaches, which would extend a few rather specific methods of synthesis known at that time [224]. The use of DMS as a solvent for performing halide – borohydride metathesis or addition of Me$_2$S·BH$_3$ complex to metal hydrides, followed by desolvation of the products at *ca.* 140 ºC under vacuum, allowed for the preparation of a whole series of monometallic borohydrides [256]. The range includes M(BH$_4$)$_n$ compounds of alkali, alkaline earth, transition and rare earth metals obtained in rather pure form [247,257-259]. Interestingly, in the case of M = Sc the mixed-cation borohydrides, MSc(BH$_4$)$_4$, are prepared in the DMS-mediated process instead of the elusive Sc(BH$_4$)$_3$, M – alkali metal [254], *e.g.*:

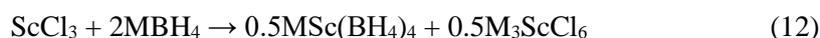

$$\text{ScCl}_3 + 2\text{MBH}_4 \rightarrow 0.5\text{MSc(BH}_4)_4 + 0.5\text{M}_3\text{ScCl}_6 \qquad (12)$$

A different approach consists of salt metathesis reactions performed in the solvents of much lower coordination ability as compared to DMS, and utilizes unique properties of weakly coordinating anions (WCA) [260-264], as in ref. [250]:

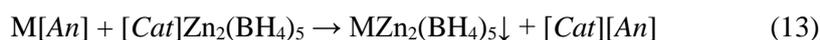

$$\text{M}[An] + [Cat]\text{Zn}_2(\text{BH}_4)_5 \rightarrow \text{MZn}_2(\text{BH}_4)_5\downarrow + [Cat][An] \qquad (13)$$

where M = Li – Cs, [*An*] – WCA, like [Al{OC(CF$_3$)$_3$}$_6$a or [B{3,5-(CF$_3$)$_2$C$_6$H$_3$}$_4$]⁻, and [*Cat*] – a bulky cation, *e.g.* tetrabutylammonium. Application of this novel method allowed for the preparation of a number of mixed-cation borohydrides based on alkali, alkaline earth, transition and rare earth metals [249-252], usually without significant content of impurities; for some of them it was not possible to achieve this without the mentioned wet approach (Fig. 7a).

The solvent-mediated method with the use of WCA enabled preparation of the metastable borohydrides, like LiY(BH$_4$)$_4$ and NaY(BH$_4$)$_4$ [251], which are not formed in mechanochemical or solid/gas reactions [231]. These compounds decompose at room temperature to the composite of corresponding MBH$_4$ and Y(BH$_4$)$_3$, however, it appeared later that heating such mixture followed by fast quenching also allows for the preparation of these elusive borohydrides [265]. This method requires certain precursors, which stimulated recent development in organic complex borohydrides, [*Cat*]*RE*(BH$_4$)$_3$, [*Cat*] = [*n*-Bu$_4$N], [Me$_4$N], [Ph$_4$N], and related salts containing magnesium and transition metals, as well as the salts of some WCA, like M[Al{OC(CF$_3$)$_3$}$_4$], M = Li – Cs, and other compounds to grow the number of accessible combinations [249,250,252,266-269]. Although the synthetic approach discussed here has been originally designed for the mixed-cation borohydrides, it can also be useful for other salts, as exemplified by M(BH$_3$NH$_2$BH$_2$NH$_2$BH$_3$), M = Li – Cs, or even Ag$_2$S$_2$O$_8$ [270,271]. It turns out that these long-chain M(BH$_3$NH$_2$BH$_2$NH$_2$BH$_3$) materials decompose thermally at rather low temperatures of 100−180 °C and respective MBH$_4$ are found in the solid residue [270,272].

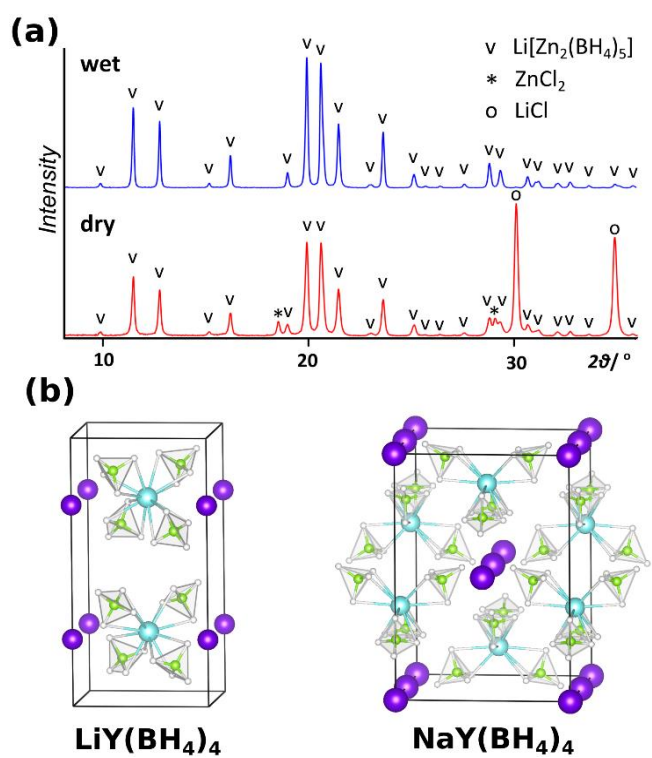

**Figure 7.**
(a) Comparison of the powder X-ray diffraction patterns of LiZn$_2$(BH$_4$)$_5$ prepared according to eq. (13) – wet, with the sample obtained in the mechanochemical procedure – dry. Note significant contamination of the latter by LiCl and ZnCl$_2$ [250], (b) The crystal structures of the metastable Li and Na yttrium borohydride accessible via the solvent-mediated approach utilizing weakly coordinating anions [251].

## 2.5.2. Properties and prospective applicability

Most of the unsolvated inorganic borohydrides of which the identity has been confirmed are crystalline materials. However, only few of them are accessible as single crystals, therefore powder diffraction remains the main technique for their structural characterization. This is often supported by the use of DFT, due to difficulties in localization of hydrogen atoms from the routine XRD data. Interestingly, neutron powder diffraction revealed that the high temperature β-phase of Y(BH$_4$)$_3$ crystallizes in $Fm\overline{3}c$ [273], while the earlier powder XRD investigations suggested the space group $Pm\overline{3}m$ with an 8 times smaller unit cell and half the lattice parameters [229]. Borohydrides reveal an impressive structural diversity, ranging from salt-like ionic structures, to various nets governed by more covalent interactions, which has been recently thoroughly discussed [185,186,274]. Among rare earth borohydrides, praseodymium (Pr) is the only element that forms three different monometallic polymorphs with different symmetries, i.e. $Pa\overline{3}$, $Fm\overline{3}c$ and $R\overline{3}c$, with the last one even having large voids and being considered as porous [253,275]. The other popular analytical techniques include vibrational spectroscopy (FT-IR and Raman), which often helps to deduce the coordination mode of borohydride anion and identify the (organic) impurities [276,277]. While nuclear magnetic resonance (NMR) utilizing various nuclides, besides identification [197,227,233,278-280] allows for evaluation of dynamic phenomena, like reorientation motions [281-283].

Magnetism of borohydrides of lanthanides (Ln) has received significant attention only very recently [244,284,285]. The borohydride phases containing Ln···HBH···Ln bridges propagating in 3 dimensions of the crystal lattice, *i.e.* α- and β-Ln(BH$_4$)$_3$, where Ln = Gd, Tb, Dy, Ho, Er, Tm, as well as selected mixed-cation borohydrides composed of isolated Ln$^{3+}$ ions forming [Ln(BH$_4$)$_4$]$^-$ complexes, namely LiYb(BH$_4$)$_4$, NaYb(BH$_4$)$_4$, KHo(BH$_4$)$_4$, RbTm(BH$_4$)$_4$ have been systematically studied using SQUID magnetometry [244]. Among these compounds α- and β-Dy(BH$_4$)$_3$, as well as NaYb(BH$_4$)$_4$ reveal weak ferromagnetic interactions, while weak antiferromagnetic interactions were observed for the other materials studied. This indicates that although the borohydride anion may serve as a transmitter of magnetic superexchange (*via* H and B atoms), the corresponding superexchange constants remain small.

As the thermolysis remains the method of choice for the release of hydrogen stored in these chemical compounds, the thermal decomposition of borohydrides has been thoroughly studied. Although the alternative hydrolytic approach has also been investigated for a few systems, it is not preferred due to high thermodynamic stability of the products of hydrolysis which would make *in situ* hydrogen re-charging an energy expensive and rather difficult process [286-288]. However, it is worth to mention that a very recent study reports significantly improved method of the reformation of NaBH$_4$ after hydrolysis [289]. While the temperature of hydrogen release from most of the known borohydrides remains too high, the other emit significant amount of diborane. Therefore, a number of means has been tested to achieve the necessary improvement of these properties. Typical approaches involve modification of chemical composition, or significant lowering of particle size and nanoconfinement [186,187]. However, in the case of yttrium borohydride it has been observed that the initial stage of thermal decomposition is facilitated for β-Y(BH$_4$)$_3$ polymorph in comparison with the more common α-Y(BH$_4$)$_3$. This manifests by lowering the temperature of the DSC event related to the first stage of decomposition by *ca.* 8 °C, and more impressively, by almost 3-fold drop of the corresponding apparent activation energy calculated using the Kissinger method, which indicates significant improvement of kinetics [290]. Although this finding demonstrated one of the possible tuning options for hydrogen release from borohydrides, its utilization would rather be limited due to additional difficulty imposed by the requirement of recreation of a particular polymorph during rehydrogenation of the spent hydrogen store (below 400 °C both polymorphs release 7.0–7.4 wt% of hydrogen, as calculated for pure Y(BH$_4$)$_3$).

Another broadly explored attempt to facilitate hydrogen release is the formation of protonic-hydridic compounds [291] *via* a combination of the moieties containing partially negatively charged H atoms bound to boron with those attached to nitrogen, which are partially positive. Such approach often increases the theoretical gravimetric content of hydrogen and lowers the temperature of decomposition, as compared with similar borohydrides lacking nitrogen [292]. Ammonium borohydride, NH$_4$BH$_4$, containing 24.5 wt.% of hydrogen and releasing most of it below 160 °C is a striking example here [293]. As the latter compound is unstable at room temperature its stabilization has been attempted *i.e.* by formation of mixed-cation borohydrides, which should render the partial charges on hydridic and protonic hydrogen atoms less pronounced:

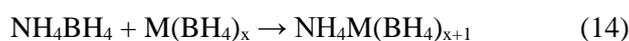

$$\text{NH}_4\text{BH}_4 + \text{M(BH}_4)_x \rightarrow \text{NH}_4\text{M(BH}_4)_{x+1} \qquad (14)$$

Such chemical stabilization of NH$_4$BH$_4$ appeared successful for NH$_4$Ca(BH$_4$)$_3$, which is stable at room temperature, and releases *ca.* 17.5 wt% of nearly pure hydrogen in several endothermic (!) steps within 95–400 °C interval [294,295]. An explanation of the unexpected endothermic character of decomposition is definitely needed [296]. A completely different behavior has been observed for the series of related borohydrides of trivalent metals: NH$_4$M(BH$_4$)$_4$, where M = Al, Sc, Y [233,297]. The latter compounds start to decompose at a temperature range comparable with the parent NH$_4$BH$_4$ (53 °C) with most of the decomposition stages being exothermic, making hydrogen release an irreversible process. The latter is rather common feature for protonic-hydridic hydrogen stores, observed even for the species containing more stable [B$_{12}$H$_{12}$]$^{2-}$ anions [298].

As it has been mentioned above, the borohydrides decomposing within the favorable temperature range of *ca.* 100 °C emit significant amount of diborane together with hydrogen [186]. Therefore, the application of a catalyst, which would rapidly and selectively decompose B$_2$H$_6$, leaving boron in the sample and purifying the released hydrogen, should greatly improve the performance of such systems. This should also be expected for much more stable borohydrides.

Recently, metallic vanadium, which forms a divalent hydride of rather moderate stability (T$_{dec.}$ of *ca.* 35 °C) [31], as well as its nanoparticles and oxides: V$_2$O$_3$, VO$_2$ and V$_2$O$_5$, have been tested as borohydride dopants [299]. Among these systems no significant influence has been observed for LiBH$_4$, and partial decomposition without further catalytic effect occurred for $Mg(BH_4)_2$. However, $NaBH_4$ doped with 25 wt% $V_2O_5$ revealed significant improvement in comparison with the parent borohydride. This catalyzed system releases 3.6 – 5.0 wt.% of H already around 380 – 390 °C, i.e. *ca.* 130 °C lower than pure $NaBH_4$ [186]. At the same time, the system doped with vanadium nanoparticles does not reveal similar effects, probably due to the ease of their surface degradation/oxidation. All the vanadium-based catalysts have also some influence on thermal decomposition of $LiZn_2(BH_4)_5$, lowering the temperature of the fastest stage of decomposition (by up to 25 °C) already on 5 wt.% doping level. However, the latter systems need more thorough research for better understanding.

Thermolysis of borohydrides can potentially be useful for the preparation of reagents or solid materials. For example, heating of $LiZn_2(BH_4)_5$ above 85 °C makes it a convenient and relatively safe source of $B_2H_6$ which can be used on-demand for different reaction processes [300]. As borohydrides decompose at temperatures of up to a few hundred °C, which is significantly lower than the temperature range typically used for the preparation of borides (reaching even 2000 °C), a few borohydrides have been assessed as precursors towards the binary compounds of boron. Simple $Mg(BH_4)_2$ and two of its solvates, $Mg(BH_4)_2 \cdot 3THF$ and $Mg(BH_4)_2 \cdot 1.5DME$, were tested for synthesis of $MgB_2$ superconductor in mostly amorphous forms [249,301]. Similarly, thermolysis of a series of lanthanide borohydrides led to their refractory borides, like $ErB_4$, $TmB_4$, $TbB_4$, and also $TmB_6$ and $TbB_6$. However the purification of these compounds appeared unsuccessful [243]. In contrast, pure amorphous quasi-hexagonal boron nitride containing minor amount of quasi-cubic form has been recently obtained *via* thermal decomposition of $(NH_4)_3Mg(BH_4)_5$ [292]. The latter compound has an exceptionally high content of hydrogen (21 wt.%). It decomposes already in the temperature range of 220 – 250 °C, while further heating up to 650 °C is beneficial for the quality and yield of obtained BN. This is significantly lower temperature than 900 – 1500 °C required in the previously known – and industrially applied – approaches towards BN [302]. Interestingly, boron nitride is the only chemical compound detected in the sample after rinsing with water, while formation of $MgB_2$ – a typical product of $Mg(BH_4)_2$ pyrolysis – is not observed in this case.

The decomposition temperatures and pathways of the aforementioned rare-earth borohydrides may depend on their synthesis method: solvent free synthesis often uses an excess ratio of 6:1 of the $LiBH_4$ to the RE halide to form a composite [303]. In such conditions partial Cl⁻ substitution occurs very often, which leads to the formation of $Li(BH_4)_{1-x}Cl_x$ with an inherited lowering of the phase transition temperature from *o*-$LiBH_4$ to *h*-$LiBH_4$ [304,305]. Moreover, selected compounds melt at 200 °C [273], and others form an X-ray amorphous material without showing signs of this aforementioned melting behavior [257]. All these solvent free composites show decomposition between $200 < T_{dec} < 300$ °C [273,306,307]. Among the solid-state decomposition products, the respective *RE* hydrides, and *RE* borides form at relatively low temperatures. LiH is also a reaction product, which seems responsible for the possibility to rehydrogenate and reform the precursor $LiBH_4$ [67,253,308,309].

# 3. Metal hydride thin films for hydrogen sensors

Efficient, reliable, and fast detection of hydrogen is a prerequisite for a successful sustainable hydrogen economy and crucial in many industrial processes and bioanalytical applications. For safety reasons, hydrogen leaks have to be detected immediately, ideally well before its concentration gets close to the safe limit of 4% in air. In addition, the monitoring of the hydrogen concentrations and pressures is vital for the efficient operation of hydrogen fuel cells, $CO_2$ conversion devices, and in a variety of industrial processes. Although there are many types of hydrogen sensors commercially available, including catalytic resistor detectors, electrochemical devices, and sensors based on changes in the thermal conductivity, all of them have major disadvantages. Apart from practical and economic drawbacks such as their relatively large size, need for regular calibration, narrow sensing range, and high costs, there are some serious safety concerns with these conventional sensors: they often require the presence of oxygen and/or current leads, forming a potential explosion hazard [25,310].

Metal hydride based hydrogen sensors are considered to be a promising alternative to these conventional hydrogen sensors. These sensors utilize the propensity of metal hydrides to hydrogenate when exposed to a partial hydrogen pressure, which, in turn, results in volumetric expansion of the metal hydride and a change in its optical and electronic properties. By probing one of these properties, the hydrogen pressure in the environment of the metal hydride sensors can thus be determined. As the hydrogenation of a metal at a given pressure typically decreases with temperature, metal-hydride based hydrogen sensors should either operate at a fixed temperature or incorporate a thermometer to account for the temperature-dependent signal. One of the major challenges for (metal hydride) hydrogen sensors is the resistance to chemical species other than hydrogen. Especially the catalytic surface of palladium is prone to poisoning by aggressive chemical species as $NO_x$ and CO as well as humidity, rendering the sensor inactive. To tackle this problem, several polymeric coatings (≈30 nm) have been developed. Notable examples include polytetrafluoroethylene (PTFE) that reduces the poisonous effect of humid air as well as poly(methylmethacrylate) (PMMA) that prevents protection against CO [26,311]. These coatings have also been shown to reduce the activation energy of hydrogen (de)sorption and as such substantially improve the response times of metal hydride hydrogen sensors. In addition, thin platinum top layers [312] or alloying palladium catalyst with elements as Cu have also been employed to reduce the poisoning effects of e.g. CO [313].

## 3.1. Sensor design and read-out

The earliest application of metal hydrides in hydrogen sensors was by probing the electrical resistivity of palladium [314]. Like most metal hydrides, the exposure of palladium to a hydrogen environment results in the hydrogenation of palladium and, in turn, to an increase of its electrical resistivity. This change in resistivity can easily be detected by a four-points resistivity measurement of a continuous thin film (< 100 nm). Alternatively, when higher sensitivities are required, the resistivity of nanofabricated textures such as patterned nanosheets, single and multiple nanowires in combination with inter-digitated transducers (IDT) and microelectromechanical-systems (MEMS) can be used.

Another class of hydrogen sensors uses an optical read-out, which have, as compared to resistive hydrogen sensors, an inherent safety benefit as they do not require any electrical currents near the sensing area (see, e.g., Silva *et al.* [27], Yang *et al.* [315], and Zhang *et al.*[28] for recent reviews). Furthermore, those sensors are typically less prone to electromagnetic radiation disturbing the readout. Several optical detection principles have been suggested that can roughly be categorized into two groups: intensity modulated and frequency modulated optical hydrogen sensors. Intensity modulated sensors are usually based on measuring the changes in the optical reflectance or transmittance of a metal hydride upon its exposure to a hydrogen environment. For example, in a micro-mirror configuration, light is coupled into an optical fibre and partially reflected by a thin film metal hydride with a thickness of 20 - 100 nm located at the tip of the fibre. The intensity of the reflected light, which alters when the metal hydride is exposed to hydrogen, is subsequently monitored by e.g. a photodiode and is directly related to the hydrogen pressure.

Frequency-modulated optical hydrogen sensors have the advantage that the read-out is insensitive to intensity fluctuations of the light source. One of the most attractive options is based on the (localized) surface plasmon resonance ((L)SPR) of a metal-hydride that may occur when it is illuminated by light that matches the frequency of the plasmon resonances. As illustrated in Fig. 8, when exposed to hydrogen, the resonance frequency changes, which may be detected by measuring the resonant frequency itself (or, alternatively, the intensity of light reflected by the sensing layer). While SPR effects occurring at the interface between a metallic surface and another dielectric medium have been considered for a longer time [316,317], especially hydrogen sensors based on LSPR in nanoparticles are recently studied intensively. In the latter case, when the dimensions of metal hydride nanoparticles are much smaller than the wavelength of the incident light, strong absorption and scattering of the incident light occurs when its frequency matches the resonance frequency of the collectively oscillating free electrons in the particle. As a consequence, a peak, the LSPR peak, is visible in the extinction spectrum of which the wavelength strongly depends on the size, shape and permittivity of the nanoparticle. When nanoparticles are exposed to hydrogen, they hydrogenate, causing a broadening and a red-shift of the LSPR peak that turns out to be directly correlated to the hydrogen-to-metal ratio in the particle and from the position or width the hydrogen pressure can thus be deduced [318]. The spectral position of this peak can then be measured by considering either the reflected or transmitted intensity.

Frequency-modulated optical hydrogen sensors have also been developed based on interferometry or Fibre-Bragg gratings [319,320]. In its most basic form, Fibre-Bragg sensors consist of (i) a fibre in which the core contains a grating of alternating layers of materials with different refractive indices and (ii) a cladding that is partly covered by a metal hydride. As such, the fibre only reflects light with a specific wavelength. Upon exposure to a partial hydrogen pressure, the metal hydride expands, which in turn changes the pitch of the grating in the core of the fibre and thus the characteristic wavelength that is reflected. As these sensors are based on the volumetric expansion of strained materials, their response times are relatively long, rendering them unsuitable for safety reasons and thus most applications.

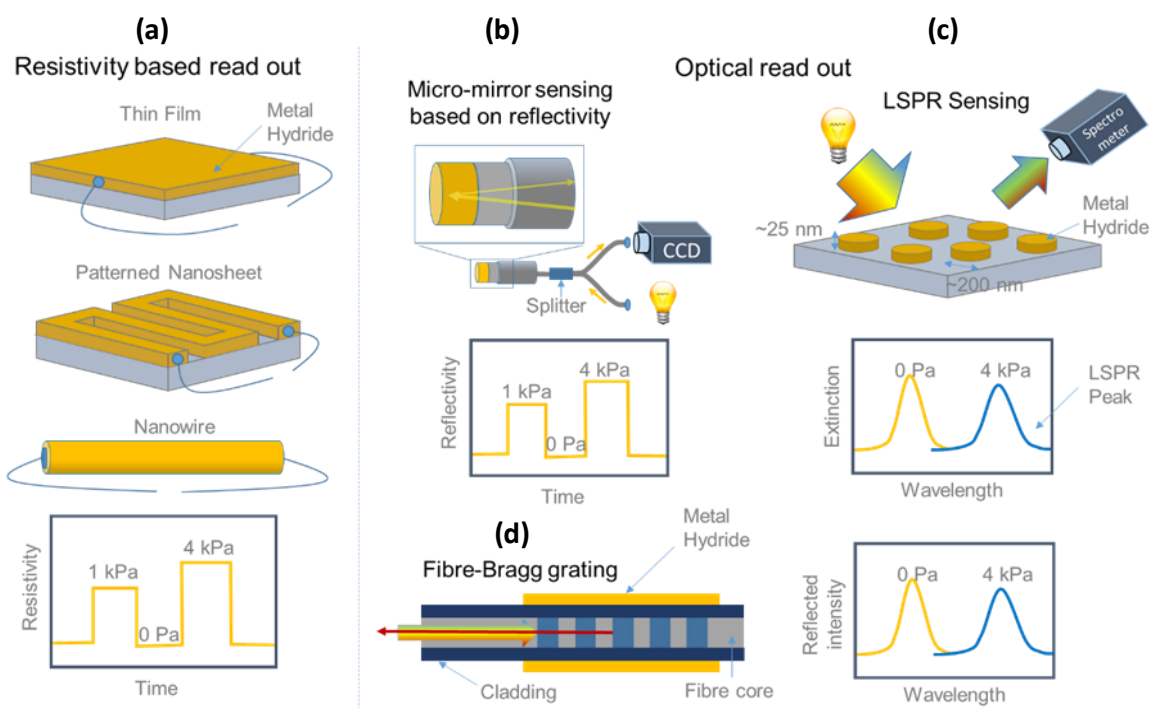

**Figure 8.**
Schematic illustration of different metal hydride based hydrogen sensors. (a) In resistivity based hydrogen sensors, the electric resistivity is measured of e.g. a metal hydride thin film, patterned nanosheet or nanowire to probe the hydrogen pressure. (b) In the micro-mirror sensor, a metal hydride sensing layer is deposited at the end of an optical fibre. Light, which is coupled into the fibre, is partly reflected by this metal hydride sensing layer, and subsequently detected by e.g. a charged-coupled device (CCD). The amount of light reflected is directly related to the hydrogen uptake of the sensing layer, and thus the partial hydrogen pressure in the environment of the sensor. (c) In sensors using the localized surface plasmon resonance (LSPR) peak, an array of nanoparticles is typically illuminated by white light, of which the transmission or reflectivity is measured. Subsequently, the extinction spectrum is computed by taking the relative difference between the incident and transmitted/reflected intensity. In the extinction spectrum, the so called LSPR peak shows up. Upon exposure to hydrogen, the LSPR peak shifts to the red, and this shift is directly related to the hydrogen pressure. (d) In Fibre-Bragg grating sensors, white light is coupled into an optical fibre in which a grating is present in its core and of which e.g. the cladding is covered with a metal hydride layer. As such, light with only one wavelength will be reflected by the grating. When exposed to hydrogen, the metal hydride layer will expand, causing a change of the periodicity of the grating, and thus of the reflected wavelength.

## 3.2. Suitable sensing materials

While there is a wide variety in the design and read-out mechanisms of metal hydride based hydrogen sensors, the requirements to the actual sensing material are largely overlapping. In order to obtain the ideal hydrogen sensor that features a large sensing range, fast response, high sensitivity, enduring stability and chemical selectivity [310], metal hydrides with a large hydrogen solubility window (preferably within one phase), no hysteresis, appropriate enthalpy and entropy change and fast hydrogen diffusion are required. In the literature, two approaches have been used to develop these hydrogen sensors: (i) materials with both a hydrogen sensing and dissociation functionality as e.g. (alloys) of palladium and (ii) materials in which the sensing and dissociation functionality are separated as e.g. palladium capped transition metals [24].

### 3.2.1. Palladium-based materials

Palladium is unarguably the most frequently considered material for metal hydride based hydrogen sensors. Its attractiveness stems from its capability to dissociate molecular hydrogen at room temperature, its modest room temperature sensing range of at least $P_{H2} = 10^{+1} - 10^{+4}$ Pa, and a reasonably fast response that can be in the order of seconds. However, palladium also has some serious shortcomings: while it already hydrogenates at pressures well before the safe limit of hydrogen in air, its sensing range is relatively limited and the mechanical stability is typically poor. The poor mechanical stability, which greatly reduces the lifetime of the sensor, has been linked to the relatively large volumetric expansion of palladium upon hydrogenation. Especially palladium thin films have been found prone to delamination from their support. However, this problem is typically easily solved by using intermediate layers of e.g. nickel [321] or titanium [322].

The most important limitation of palladium-based hydrogen sensors is the severe hysteresis in the readout: as palladium undergoes a first-order metal-to-metal hydride transition from the dilute α to the high-hydrogen concentration metal hydride β-phase, the readout depends strongly on the sensor's pressure history. Alloying of palladium has widely been employed to obtain more reproducible sensing characteristics at room temperature. Typically, elements including Au [318,323-326], Ag [327], Ta [328], Cu [313] and Ni [329,330] are introduced in palladium, tuning the d-band and shifting the critical temperature of $T_C \approx 290$ °C down, thereby suppressing the first-order phase transition for sufficiently high concentrations of these dopants [331]. Unfortunately, a small hysteresis can often still be discerned even though the first-order transition is completely eliminated at high dopant concentrations. For example, in the case of $Pd_{1-y}Au_y$ thin films display a small hysteresis over a wide pressure range owing to profound clamping of the film to the support: upon hydrogenation, the volumetric expansion of the cubic unit cell has to be completely translated into a thickness increase of the film. Such effects are not seen in nanoparticles as they have the ability to expand more freely. The different appearance of hysteresis as well as their dissimilar hydrogen solubilities emphasize that the confinement of materials at the nanoscale the way materials are nanoconfined can dramatically influence its properties and thus their sensing characteristics [332]

Alloying palladium also affects the hydrogen solubility. Introducing elements with a larger unit cell like Au, Ag, and Ta increase the hydrogen solubility at lower pressures, thus extending the sensing range. For example, in the case of thin film and nanoparticles of $Pd_{0.7}Au_{0.3}$, an extraordinary sensing range of $P_{H2} = 10^0 – 10^{+6}$ Pa has been demonstrated, which likely extends to even higher pressures making it a suitable candidate to probe relatively high hydrogen pressures. However, alloying palladium typically reduces the maximum hydrogenation of the metal hydride and thus greatly reduces the maximum change of the optical and electronic properties of the material, thereby compromising the sensitivity of the hydrogen sensor.

### 3.2.2. Materials with separated hydrogen dissociation and sensing functionality

Separating the hydrogen dissociation and sensing functionality considerably broadens the scope of metal hydride materials that can be used in a hydrogen sensor. Such sensors with separate functionalities have mainly been developed using an optical readout and consist of at least two layers: (i) a sensing layer that is responsible for the response to hydrogen, typically transition metals and metal hydrides that exhibit a metal-to-insulator transition upon hydrogenation, and (ii) a capping layer that takes care of the dissociation of hydrogen as e.g. a thin (≈ 10 nm) palladium layer.

Transition metals including hafnium and tantalum have very advantageous sensing properties [333,334]. In these materials, hysteresis arising from a first-order transition is easily circumvented by considering materials with a large solubility window within one single phase. $TaH_x$ is a notable example: it has a large solubility range of $0 < x < 0.7$. As a result, it entails an exceptionally broad sensing range of at least $P_{H2} = 10^{-2} – 10^{+4}$ Pa at $T = 120$ °C. In addition, as the hydrogen uptake of tantalum is considerable larger than for palladium alloy sensing materials, its (optical) contrast is also substantially larger, allowing the development of hydrogen sensors with a better sensitivity [333].).

The fact that materials as magnesium and yttrium show a metal-to-insulator transition upon hydrogenation implies a substantial change to the optical and electronic properties when these materials are exposed to hydrogen. This large change in properties, which have e.g. also been exploited in switchable mirrors, potentially allows for the development of hydrogen sensors with high sensitivities. However, the applicability of both materials is greatly reduced by the first-order metal-to-metal hydride transition, which implies a profound hysteresis. In addition, magnesium and, to a smaller extend yttrium, only exhibit a limited pressure range around their plateau pressures in which they are sensitive to hydrogen. Alloying as well as using the profound thickness dependence of the plateau-pressure of magnesium based films have been employed to increase the sensing range (see, e.g., [335-341]) and also facilitated the development of low-cost eye-readable hydrogen sensors [337,339,340]. Unfortunately, these materials typically form extremely stable hydrides and exhibit therefore very slow desorption kinetics, limiting their practical applicability to a one-time-use indicator of the highest hydrogen pressure that was present in the environment of the sensor.

# 4. Application in secondary batteries

## 4.1. Metal hydride electrodes for LIBs

Metal hydrides have been considered as potential anode materials for Li-ion batteries due to their large capacities and low average potential value versus $Li^+/Li^0$. Since the discovery of the electrochemical activity of metal hydrides towards lithium, $MgH_2$ (theoretical capacity ~2000 mAh g$^{-1}$) remains the most studied in this category of anodes for LIBs using a carbonate-based liquid electrolyte [17,18,342-349]. It can deliver a reversible capacity of 1480 mAh g$^{-1}$ during the total conversion reaction (eq. 15):

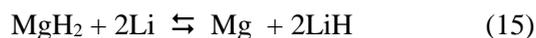

$$MgH_2 + 2Li \leftrightarrows Mg + 2LiH \qquad (15)$$

However, though the high theoretical capacity, pristine $MgH_2$ electrode suffers from large initial irreversibility and capacity fading after few cycles, which is connected to the high volume variation (~85%) and LiH interaction with the liquid electrolyte during the (dis)charge processes [343,346,350]. At low cycling rate (1Li/100h), $MgH_2$ reacts with Li showing a full discharge ($\Delta x \approx 2.9$ Li) with two plateaus at 0.44 V ($\Delta x \approx 1.8$ Li) and 0.095 V. For the first plateau, XRD shows the appearance of the *hcp* Mg-type and *bcc* Li-type solid solutions. The formation of these solid solutions can be avoided by limiting the discharge at $\Delta x \approx 2$ Li or 0.15 V using in-house prepared $MgH_2$ (Mg milled with 10% MCMB 2528 then hydrogenated under 20 bar $H_2$ at 350 °C for 10 h). A reversible capacity of 1500 mAh g$^{-1}$ (25% loss) is reached for the first step, while a reversible capacity 2700 mAh g$^{-1}$ (33% loss) is measured for both steps [17].

Electronic transfer is an important parameter during any electrochemical reaction. The poor electric conductivity of hydrides has to be taken in consideration. For instance β-$MgH_2$ and γ-$MgH_2$ insulators exhibit band gap energies of 5.6 and 5.3 eV, respectively [351]. In addition, the conductivity is also influenced during cycling by the formation of the conducting metallic Mg and insulating LiH. For example, in thin films the presence of small amounts of Mg can be sufficient to reach the minimum conductivity in the electrode, where the addition of carbon is not necessary. Owing to the metallic behavior of $TiH_2$, its addition to bulk $MgH_2$ has been also considered as composite anode for LIBs [17,18].

DFT [352,353] has been regarded as a powerful tool to calculate and predict the structures, energies and electrochemical properties of hydrides materials for conversion-type electrodes applications. Ramzan *et al.* [354] performed *ab initio* simulations to investigate the MgH$_2$ electrode, which was in good agreement with experiments in terms of the average working voltage for Li-ion battery. However, their calculation of doped-MgH$_2$ as conversion reaction anode did not show much improvement in terms of diffusion of lithium. Qian *et al.* [355,356] explored pure and various metal-doped NiTiH hydrides as anode materials for Li-ion batteries using DFT, in which the enhanced electrochemical capacity and a minor increase in voltage were found in Li-doped NiTiH. The effects of various light-metals (Mg, Al) and transition metals (V, Cr, Mn, Fe, Co, Cu, Zn) on the electrochemical properties of NiTiH conversion electrodes were also screened in detail by the authors. The doping of Al, Cr, Mn or Fe was found to have the most remarkable effects on pristine NiTiH hydride.

Through materials design and predictions using DFT, conversion electrodes of Li-ion batteries with other complex hydrides also show excellent properties. Mason *et al.* [357] determined the phase diagram in the Li-Mg-B-N-H system in the grand canonical ensemble, as a function of lithium chemical potential. Such calculations have also been conducted to study the capacities regarding several new conversion reactions. Qian *et al.* [358] explored the pure and Li-doped Mg$_2$NiH$_4$ as conversion anode materials for Li-ion battery applications using DFT and AIMD technique. The most thermodynamically stable Li-doped Mg$_2$NiH$_4$ possesses a smaller band gap than pure Mg$_2$NiH$_4$ and has a theoretical specific capacity of 975.35 mAh g$^{-1}$ and an average voltage of 0.437 V (versus Li$^+$/Li$^0$). The diffusion behavior of Li-ions in this electrode material at 300 K is also improved by Li-doping. In particular, the diffusion coefficient of Li-ion is increased evidently after Li-doping. Moreover, Mg$_2$NiH$_4$ doped with nine different elements (Na, Al, Si, K, Ca, Ti, Mn, Fe, Co) has been predicted as conversion-type electrode materials through computations [359]. The electrochemical properties such as specific capacity, volume change and average voltage as well as the electronic band gap of different doped systems have been calculated as shown in Fig. 9. The Na-doped material shows the highest electrochemical specific capacity. The dopants Si and Ti have the most obvious effects to reduce the electronic band gap of the electrode material. All nine doping elements can help to reduce the average voltage of the negative electrodes and tend to have acceptable volume changes. Pure and various doped Mg(AlH$_4$)$_2$ have also been reported to exhibit remarkable specific capacities [360]. The theoretical specific capacity of the Li-doped material can be 2547.64 mAh g$^{-1}$ with a small volume change of 3.76% during the electrode conversion reaction. The strong hybridization between Li s-state and H s-state influences the performance of the Li-doped material. It is believed that DFT modeling would further help the design and prediction of better light-metal based complex hydrides for conversion electrode applications.

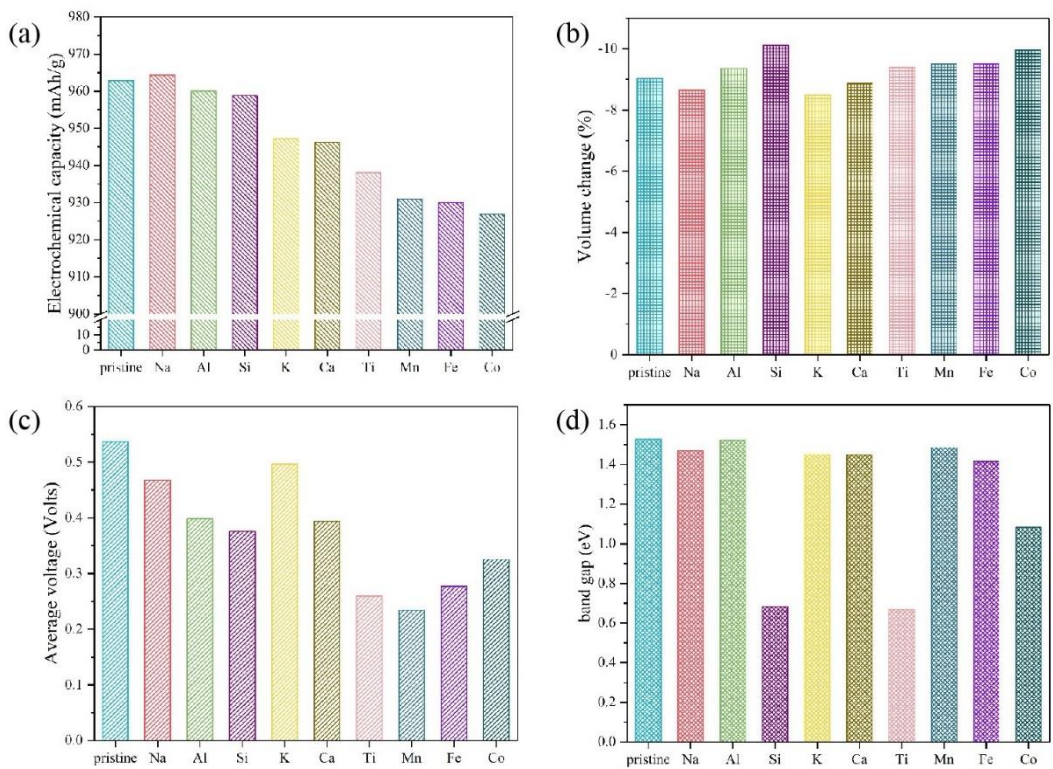

**Figure 9.**
Prediction of properties of various doped $Mg_2NiH_4$ using DFT: (a) electrochemical lithium-storage capacity; (b) volume change; (c) average voltage (versus $Li^+/Li^0$) ; (d) electronic band gap [359].

In addition to its use as a solid state electrolyte (see § 4.2.1), $LiBH_4$ was also theoretically and experimentally investigated as conversion type anode material for Li-ion batteries [361,362]. Other single-cation alkaline and alkaline earth metal borohydrides $NaBH_4$, $KBH_4$, $Mg(BH_4)_2$ and $Ca(BH_4)_2$ were simultaneously investigated. All these borohydrides have indeed low average molar masses and a quite substantial hydrogen content, resulting in outstandingly large theoretical inherent gravimetric capacities in the range of 2000–4000 mAhg$^{-1}$ [362]. For $LiBH_4$ notably, electrochemical capacity of 4992 mAh g$^{-1}$ can be expected, with the following conversion reaction:

$$LiBH_4 + 3\,Li \rightarrow B + 4LiH \qquad (16)$$

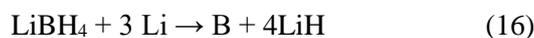

In addition, a reaction potential of 0.42 V vs. $Li^+/Li°$ associated with a favorable reaction energy of -123.3 kJ mol$^{-1}$ are theoretically predicted. For the other borohydrides, one or two steps conversion mechanisms are conceivable. In the first step, elemental boron and metal (Na, K, Mg, Ca) associated with lithium hydride are directly generated, whereas a metal hydride intermediate (NaH, KH, $MgH_2$ and $CaH_2$) is produced before the metal itself in the second step. However, from the electrochemical tests vs. $Li^+/Li°$ using $LiPF_6$/EC:DMC 1:1 (LP30) as liquid electrolyte, it appears unfortunately that $LiBH_4$ present almost no electrochemical reactivity and reversibility. In the case of other borohydrides, only Mg(BH$_4$)$_2$ and NaBH$_4$ show some first discharge capacities, but far lower as compared to the theoretical ones, only 540 and 250 mAh g$^{-1}$ respectively, and very weak reversibility. Despite these poor results, LiBH$_4$ and more generally alkaline and alkaline earth metal borohydrides, deserve to be deeply investigated mainly in composites, because of the electrochemical activity for some of them vs. Li$^+$/Li°. In that respect, with an anode composite formulation optimization going hand in hand with an electrode structure control, performances could be greatly enhanced. As a proof of concept, this enhancement was observed for MgH$_2$ as a conversion material vs. Li$^+$/Li°, with no electroactivity observed for the pristine hydride without the suitable formulation [363,364]. Moreover, in the case of the Mg(BH$_4$)$_2$ conversion, MgH$_2$ could be formed during reduction and subsequently reacts in the second step, which is conceptually interesting, even if Mg(BH$_4$)$_2$ conversion is not reversible.

## 4.2. Borohydride-based electrolytes for ASSBs

### 4.2.1. LiBH$_4$-based electrolytes

The complex hydride LiBH$_4$ and its composites (e.g. with oxides, halides, sulfides) are of considerable interest for solid-state battery applications [365,366]. In particular as solid-state electrolyte, the ionic conductivity of LiBH$_4$ is greater than 10$^{-3}$ S cm$^{-1}$ at a temperature of 393 K [367]. Therefore, an important question, from modelling point of view, is whether the high Li ion mobility in LiBH$_4$ can be extended to room temperature or if other complex hydrides with this property are available. It has been proven that mixtures of metal borohydrides with halides can stabilize the *h*-LiBH$_4$ phase at lower temperatures. This has led several authors, for example Yin *et al*. [368] and Corno *et al*. [369] to investigate anion substitution by F in the *o*-LiBH$_4$ (space group *Pnma*) [370]. Moreover, Maekawa *et al*. [192] proposed a superionic solid solution phase at room temperature by doping with lithium halides and also demonstrated that Li(BH$_4$)$_{0.667}$Br$_{0.333}$ conserves the hexagonal structure (space group *P*6$_3$*mc*) from 293 to 573 K [371]. Further, Gulino *et al.* [372] studied the mixed halide anions (Cl/Br) substitution resulting in additional optimization of the electrochemical properties of the hexagonal phase. The influence of different halide anions on the hexagonal structure is studied using DFT modelling, mainly with respect to the following properties:

- The structural and electronic changes during the substitution in the structure Li (BH$_4$)$_{1-x}$M$_x$ (with M = F, Cl, I, Br and x = 0.125, 0.25, 0.375)
- The mechanical stability of the *h*-LiBH$_4$ phase during substitution by halide anions based on atomic vibrations

Atomic relaxation and phonon dispersion have been performed for each phase in this study as preliminary results. In the high temperature phase, the [BH$_4$]$^-$ anions are arranged along the z-direction in the hexagonal plane, with two nearly equivalent Li$^+$ ion sites (Fig. 10). The obtained equilibrium volume, lattice constant and interatomic distances are summarized in Table 3. It can be noticed that LiB, BH, H-H and LiH distances are almost similar for all systems. The structural data agree well with the reported ones for chlorine substitutions at close compositions [305]. Furthermore, the interatomic distance between Li$^+$ and the nearest added halide anions varied from 1.80 to 2.77 Å. Note that, the transition *Pnma*→*P*6$_3$*mc* induces a volume expansion of 50% per formula unit. Nevertheless, for the substituted structures, it has been demonstrated that there is a significant reduction in volume, owing to a large contraction along the *a* and *c*-axis. The substitution of [BH$_4$]$^-$ by halide anions (F$^-$, Cl$^-$, Br$^-$, I$^-$) has also been investigated considering three compositions (molar fraction), x equal to 0.125, 0.25 and 0.375 in the hexagonal framework. As a contribution to this review work, our study aims to identify the microscopic behavior during the replacement of the [BH$_4$]$^-$ anion by halide (F$^-$, Cl$^-$, Br$^-$ or I$^-$) into the *h*-LiBH$_4$ phase.

The effect of halides substituted in *h*-LiBH$_4$ can be explained based on the interatomic distance. As the size of the halide ion is larger, the interatomic distance during the substitution is reduced. According to Fig. 10, a reorientation of [BH$_4$]$^-$ is observed when [BH$_4$]$^-$ anions are randomly replaced by halide ions (F$^-$, Cl$^-$, Br$^-$ or I$^-$) that becomes more pronounced with increasing substituent fraction x. While for low values of x a partial reorientation of the nearest [BH$_4$]$^-$ anion to the halide anion occurs, total reorientation of [BH$_4$]$^-$ anion to the halide anion occurs for x = 0.25 and 0.375. Differently, for x = 0.50, the orientation [BH$_4$]$^-$ is again along the z-axis and the structure becomes unstable.

The phonon dispersion has been studied to analyze the structural stability of LiBH$_4$ substituted by halides. Fig. 10 (d-g) shows that phonon modes are imaginary in case of the substitution for x = 0.125 and lower than observed for *h*-LiBH$_4$. Furthermore, the intensity of the imaginary modes decreases when the halide anion fraction is increased. As a result, the disordered phase is favored, and the phase is more mechanically stable at low temperature, due to the reorientation disordering of anion which is a common behavior of [B$_n$H$_n$]$^-$ groups as described for different systems based on borohydrides [373-375].

Table 3. Summary of DFT data results for lattice parameter, interatomic distances, optimized equilibrium cell volume and total energy of $o$-LiBH$_4$, $h$-LiBH$_4$ and Li(BH$_4$)$_{1-x}$X$_x$, (with X= F, Cl, Br, I and x= 0.125, 0.25, 0.375) in hexagonal phase, respectively.

| System | Parameters (Å) | | | Interatomic distances (Å) | | | | | Volume (Å$^3$/f.u) | Total energy (Ry/f.u) |
|---|---|---|---|---|---|---|---|---|---|---|
| | a | b | c | Li-H | B-H | H-H | Li-B | Li-X | | |
| $o$-LiBH$_4$ | 7.21 | 4.36 | 6.61 | 1.98 | 1.22 | 1.97 | 2.47 | | 206.01 | -25.98 |
| $h$-LiBH$_4$ | 4.19 | 4.19 | 7.30 | 1.89 | 1.22 | 1.99 | 2.48 | | 106.06 | -25.97 |
| $h$-LiBH$_4$ (supercell 2x2x1) | | | | | | | | | | |
| Li(BH$_4$)$_{0.875}$F$_{0.125}$ | 4.08 | 4.08 | 6.25 | 1.90 | 1.22 | 1.96 | 2.46 | 2.03 | 90.40 | -35.43 |
| Li(BH$_4$)$_{0.75}$F$_{0.25}$ | 4.08 | 4.08 | 6.120 | 1.92 | 1.22 | 1.99 | 2.48 | 1.94 | 84.24 | -40.17 |
| Li(BH$_4$)$_{0.625}$F$_{0.375}$ | 4.01 | 4.01 | 5.84 | 1.93 | 1.22 | 1.99 | 2.50 | 1.80 | 79.73 | -28.47 |
| Li(BH$_4$)$_{0.875}$Cl$_{0.125}$ | 4.12 | 4.12 | 6.62 | 1.87 | 1.22 | 1.95 | 2.46 | 2.47 | 48.57 | -30.97 |
| Li(BH$_4$)$_{0.75}$Cl$_{0.25}$ | 4.13 | 4.13 | 6.47 | 1.91 | 1.22 | 1.97 | 2.46 | 2.45 | 46.65 | -33.47 |
| Li(BH$_4$)$_{0.625}$Cl$_{0.375}$ | 4.10 | 4.10 | 6.40 | 1.92 | 1.22 | 1.98 | 2.46 | 2.42 | 45.59 | -28.78 |
| Li(BH$_4$)$_{0.875}$Br$_{0.125}$ | 4.14 | 4.14 | 6.66 | 1.90 | 1.22 | 1.96 | 2.46 | 2.60 | 99.11 | -30.71 |
| Li(BH$_4$)$_{0.75}$Br$_{0.25}$ | 4.11 | 4.11 | 6.63 | 1.87 | 1.22 | 1.96 | 2.45 | 2.60 | 98.16 | -33.08 |
| Li(BH$_4$)$_{0.625}$Br$_{0.375}$ | 4.17 | 4.10 | 6.36 | 1.91 | 1.22 | 1.97 | 2.45 | 2.65 | 98.08 | -25.98 |
| Li(BH$_4$)$_{0.875}$I$_{0.125}$ | 4.17 | 4.17 | 6.75 | 1.88 | 1.22 | 1.99 | 2.48 | 2.77 | 101.53 | -31.58 |
| Li(BH$_4$)$_{0.75}$I$_{0.25}$ | 4.23 | 4.23 | 6.75 | 1.89 | 1.22 | 1.97 | 2.48 | 2.72 | 104.31 | -34.38 |
| Li(BH$_4$)$_{0.625}$I$_{0.375}$ | 4.28 | 4.28 | 6.82 | 1.90 | 1.22 | 1.96 | 2.48 | 2.73 | 107.99 | -28.34 |

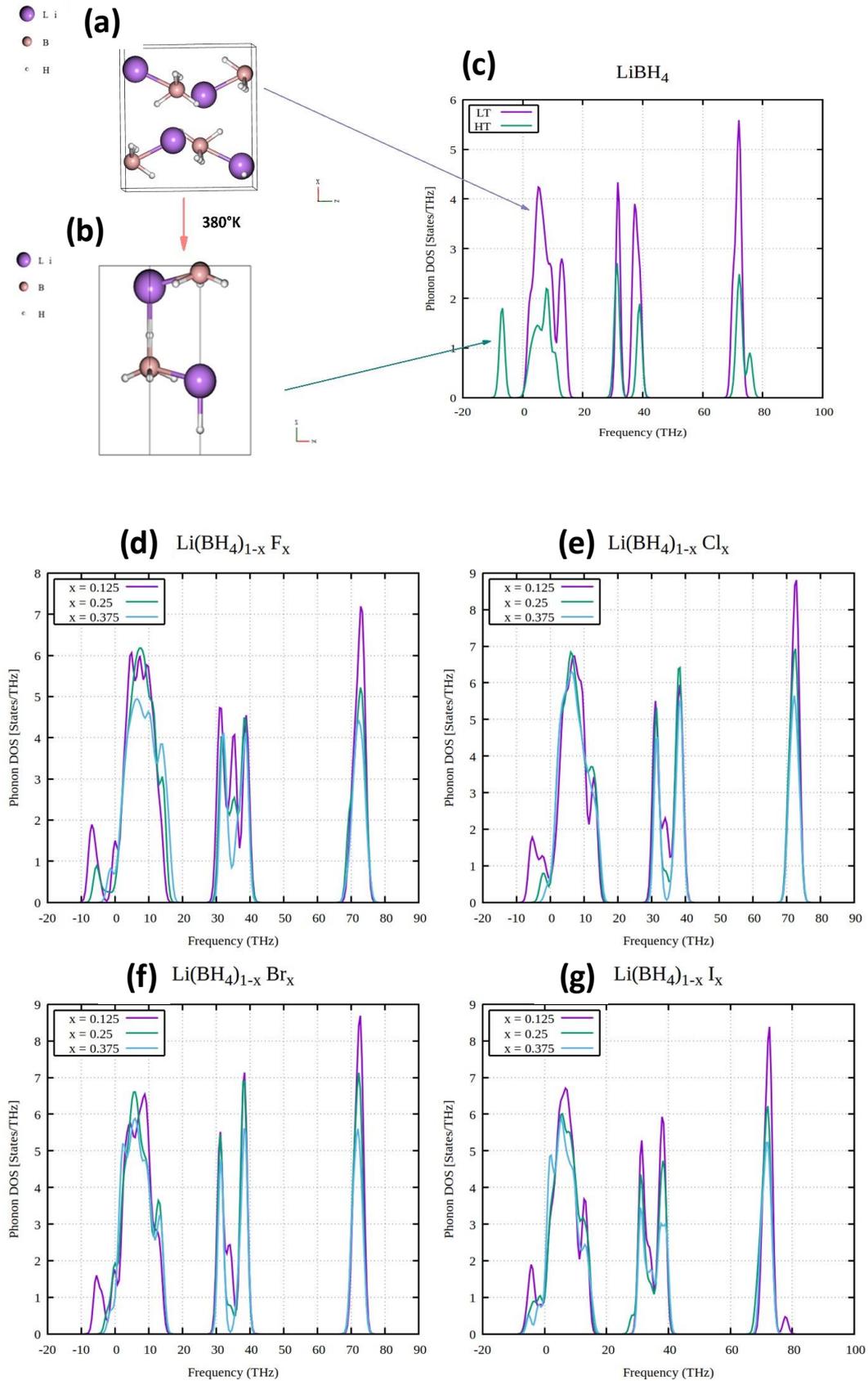

**Figure 10.**
Atomic structures of LiBH$_4$ (a) at low temperature (*o*-LiBH$_4$) and (b) high temperature (*h*-LiBH$_4$); (c)-(g) Phonon density of states of the *h*-LiBH$_4$ and *o*-LiBH$_4$ phases, and of Li(BH$_4$)$_{1-x}$X$_x$ (with X= F, Cl, Br, I and x= 0.125, 0.25, 0.375).

With respect to application in batteries, Li(BH$_4$)$_{0.75}$I$_{0.25}$ solid electrolyte has been studied and revealed to be suitable for application in LIBs [376,377]. The remarkable properties of LiBH$_4$-based materials reside in their reducing character, ductility and compatibility with Li metal anode, which make them as potential candidates for solid-state electrolytes. Progress has been made to improve the ionic conductivity of LiBH$_4$-based systems at RT. Similar to halide substitution, the presence of (PS$_4$)$^{3-}$ groups embedded in the structural unit of Li(BH$_4$)$_{0.75}$I$_{0.25}$ could allow the accommodation of the BH$_4$-BH$_4$ distances and less hindered effect for Li mobility [350,378]. These composite electrolytes exhibit high ionic conductivity, ~ 10$^{-3}$ S cm$^{-1}$, and good chemical compatibility compared to pure LiBH$_4$, as well as an electrochemical window up to 5 V. Unlike the Li(BH$_4$)$_{0.75}$I$_{0.25}$ system, Li(BH$_4$)$_{1-x}$Cl$_x$ reverts back to the initial phases, implying the formation of LiBH$_4$ and LiCl upon cooling to RT [370,379]. Hauback and co-workers studied further the pseudo-ternary LiBH$_4$-LiCl-P$_2$S$_5$ system (Fig. 11a) and showed possible incorporation of Cl$^-$ in the structure when small amounts of P$_2$S$_5$ are added.

Using the electrolyte composite (LiBH$_4$)$_{0.73}$·(LiCl)$_{0.24}$·(P$_2$S$_5$)$_{0.03}$, a battery-cell is demonstrated in the presence of TiS$_2$ electrode and Li metal anode (Fig. 11b). Furthermore, to monitor the (de)lithiation processes an *operando* SR-XRD has been demonstrated. Thanks to the low scattering of the bulk electrolyte (Fig. 11c), the expansion of TiS$_2$ can be explicitly observed which occurs only in *c* direction due to the layered structure of the material [380], and no changes in the bulk electrolyte itself can be detected. The *operando* SR-XRD setup in transmission mode points to a mean of fast and high resolution structural characterization of a borohydride-based solid-state battery.

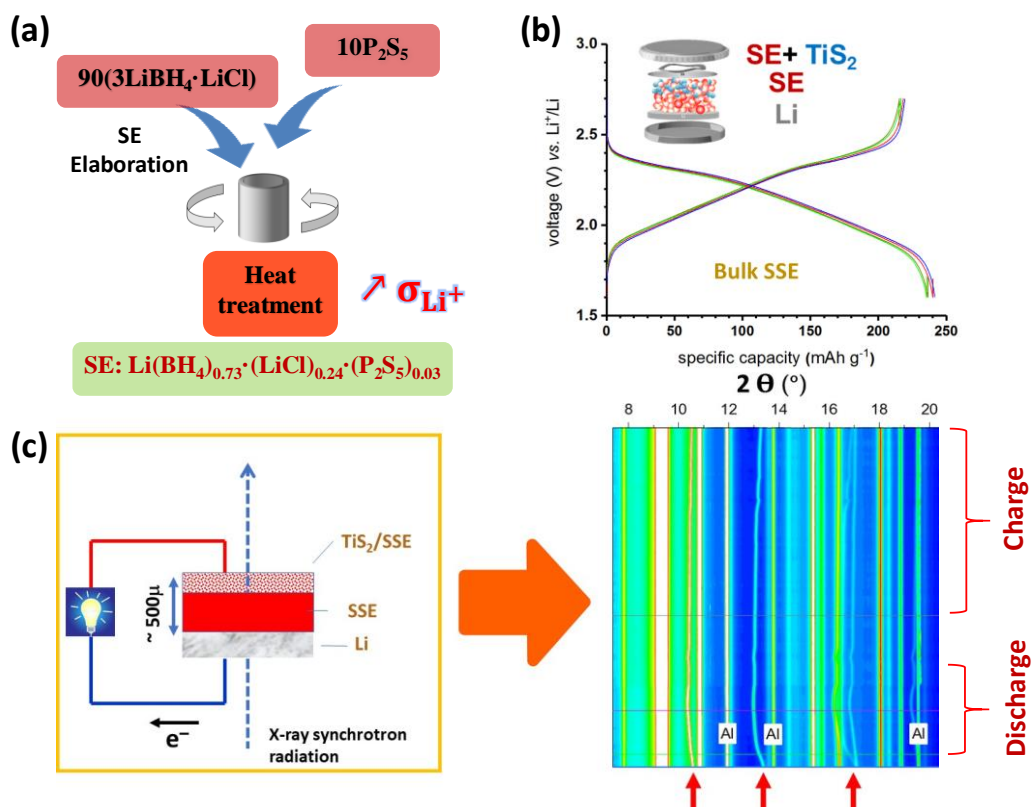

**Figure 11.**
(a) Preparation procedure of the solid electrolyte (SE); (b) galvanostatic discharge/charge cycling (0.05 C-rate) at 50 °C with TiS$_2$ electrode; (c) sketch of the experimental set-up (left) and 2D *operando* SR-PXD patterns (right) of the solid-state battery cell during discharge/charge (partial (de)lithiation of TiS$_2$ electrode, C/10, 60 °C). Red arrows indicate the reflections with the main changes [380].

### 4.2.2. Mg(BH$_4$)$_2$: dynamics, structure and ion conduction

Current materials considered for all-solid-state batteries (ASSBs) include Na$^+$, Mg$^{2+}$ and Ca$^{2+}$ –based compounds, while Al$^{3+}$–based compounds are often marginalized due to the lack of suitable electrode materials. This section aims to review the dynamics, structure and ionic conduction process for novel solid–state electrolytes based on complex metal hydrides containing magnesium (Mg) and boron (B), as well as selected solvated derivatives (e.g. ethylenediamine, diglyme and ammonia). The complex Mg(BH$_4$)$_2$ in addition to the aforementioned solvated derivatives is known to form stable solids, which are very promising Mg ion conductors. Other chemistries for operation in ASSBs are discussed elsewhere [274,381,382].

Rechargeable magnesium batteries are advantageous in many regards when compared to Li-ion battery technology [383]. The low electrochemical potential of –2.4 V (Mg/Mg$^{2+}$) *versus* a standard hydrogen electrode (SHE) is slightly higher than the electrochemical potential of –3.0 V of Li/Li$^+$. However, the natural abundancy of Mg in the earth crust is more than 2% as compared to 0.0065% for lithium [384] and 0.001% of boron [385], making Mg not only beneficial from a cost-perspective but as its mining may involve also less geopolitical challenges that may arise from the water intensive production of Li in South America [386]. Furthermore, Mg is non–toxic, easily manipulated and can be handled in air [387,388]. Another advantage for Mg is its non–dendritic formation during plating on the metal [389]. This is a key advantage over Li, in particular for ASSBs, as dendrite growth is one of the main failure mechanisms in ASSBs based on Li, although a recent study suggests a new coating to avoid dendrite formation in lithium ASSBs [390]. Consequently, 'pure' Mg metal, with a decent volumetric capacity of 3833 mA·cm$^{-3}$, could be used as safe and reliable anode material, although, the appearance of Mg dendrites was suggested recently [391]. The benefits of bivalent Mg–ion migration is also still debated in literature and it has also hampered developments of suitable cathode materials [381].

Since the discovery of Mg batteries by Aurbach *et al.* [392] in 2000, research in this field has expanded enormously [381] and even borohydrides were found to have significant conductivities of the metallic cations. With the discovery of high-ionic conductivity in hexagonal-LiBH$_4$ at ~115 °C [393], much effort has been placed into the research of borohydrides for solid-state electrolytes [274]. The first study of Mg ion conductivity in β-Mg(BH$_4$)$_2$ was by Matsuo *et al.* by first-principles molecular dynamics simulations [394]. They determined that due to the close distance of the [BH$_4$]$^-$ anions to Mg, the Mg-ions cannot move freely. The authors suggested a partial substitution of [BH$_4$]$^-$ by AlH$_4^-$ to increase distances and open the possibility of Mg$^{2+}$ conductivity in β-Mg(BH$_4$)$_2$. However, since then, the approaches of forming electrolytes can be roughly divided into liquid and solid-state fractions. The first liquid electrolyte was developed by Mohtadi *et al.*, whose study demonstrated the possibility to employ Mg(BH$_4$)$_2$ dissolved in THF and DME in a secondary magnesium battery [395]. Subsequently, Mg(BH$_4$)$_2$ was more and more replaced, by Mg(BR$_4$)$_2$ in DME (R=–OCH(CF$_3$)$_2$). So far, this complex shows the highest reported electrochemical stability window of 4.3 V and is stable in air with a Mg-ion conductivity of σ = 0.011 S·cm$^{-1}$ in a solution of 0.3 M DME [396].

The second category of solid-state Mg-ion conductors was introduced by Roedern *et al.* [397]. They synthesized a new compound from Mg(BH$_4$)$_2$ and ethylenediamine (C$_2$H$_8$N$_2$, 'en'), which was reported to have a Mg-ion conductivity of up to σ = 6·10$^{-5}$ S·cm$^{-1}$ at 343 K [397]. The synthesis was merely based on ball milling techniques and thermal treatment, while the structure of Mg(en)$_1$(BH$_4$)$_2$ has not been reported so far. This work was expanded to further solvated derivatives, such as diglyme, with measured conductivities of σ = 2·10$^{-5}$ S·cm$^{-1}$ at 350 K of Mg(BH$_4$)$_2$-diglyme$_{0.5}$ [398]. Amides also play its role among these conductors and in 2014, Mg(BH$_4$)(NH$_2$) was shown to be a solid-state Mg ion conductor [399]. More recently, a conductivity of σ = 3·10$^{-6}$ S·cm$^{-1}$ at 373 K was reported for (Mg-B-N-H)-system (of unknown stoichiometry) based on Mg(BH$_4$)(NH$_2$) and a correlation to an amorphous phase was postulated [400]. A very recent study even suggests a conductivity of σ =3.3·10$^{-4}$ S cm$^{-1}$ at T = 353 K for Mg(BH$_4$)$_2$·NH$_3$ as well as a new approach to explain the conductivity in these materials [401]. Most of the aforementioned authors reported that some amorphous phase, possibly amorphous Mg(BH$_4$)$_2$, has a beneficial influence on the conductivities. This fact is not unheard of and amorphization has also been mentioned to be advantageous for higher ionic conductivity in material classes based on Li$_3$PS$_4$, which can be identified as glassy solid-state electrolytes [402,403]. Therefore, it seemed helpful to investigate the influence of this amorphous phase on the conduction process in complex borohydrides combined with solvated derivatives. Recently, amorphous-Mg(BH$_4$)$_2$ has been investigated [215] by QENS, EIS and PDF analysis. The QENS data depicted as the scattering function S(Q, ΔE) are shown in Fig. 12. These data were used to explore the differences in internal dynamics between the crystalline γ-phase (blue circles) and the amorphous Mg(BH$_4$)$_2$–phase (red squares) at 310 K. In general, QENS is used to analyse dynamic processes such as diffusion or jump rotations – mainly of hydrogen containing species due to the high incoherent cross section of hydrogen. These motions are visible via a broadening around the elastic line which is defined at an energy transfer ΔE = 0 meV.

The data depicted in Fig. 12 show that the quasi-elastic and the inelastic contribution are strongly reliant on the local structure. The γ-phase exhibits hardly any stochastic motion or quasi-elastic contribution around the elastic line at an energy transfer ΔE=0 meV. In contrast, the amorphous phase reveals a significant broadening around this elastic line, which is indicative of higher rotational mobility of the [BH$_4$] moieties [215].

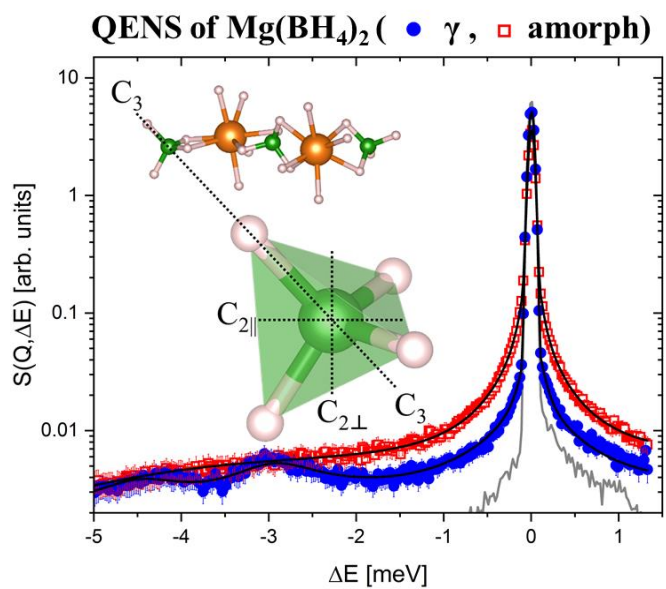

**Figure 12.**
Quasi-elastic neutron scattering (QENS) studies of Mg(BH$_4$)$_2$ in crystalline γ-modification (blue circles), and amorphous modification (red squares). S(Q, ΔE) measured at λ$_1$= 5 Å, Q = 1.35Å$^{-1}$ at T = 310 K. The solid grey curve shows the measured resolution function at 3.5 K. The solid black curves represent the fit to the data. Data from TOFTOF instrument, MLZ [404]. Inset is showing three [BH$_4$] tetrahedra in their respective Mg setting and a magnification into one tetrahedra and its rotational axes.

EIS data show that the conductivity of amorphous Mg(BH$_4$)$_2$ at 80 °C is almost two orders of magnitude higher compared to as received γ-Mg(BH$_4$)$_2$. As the PDF analysis already showed similar structural local building blocks, it was suggested that also the conduction pathway might be similar. However, analysis of the QENS measurements demonstrated that the number of activated [BH$_4$] rotation is significantly higher in the amorphous phase. Under the assumption of similar conduction pathways, it seems likely that those activated rotations support the movement of Mg-ions. This mechanism is already known as the paddle wheel effect [405], and these findings are supported by the fact that when the paddle wheel mechanism stops at the crystallization temperature (i.e. the rotating [BH$_4$] become inactive) the conductivity drops to the same level as crystalline γ-Mg(BH$_4$)$_2$ [215].

All structures of Mg(BH$_4$)$_2$ mentioned above consist of the same building blocks as shown in Fig. 6 (§ 2.4) and Fig. 12. The structural variety found for Mg(BH$_4$)$_2$ is caused by the ground state energies, which are almost degenerate [209,406,407]. Previous investigations of the dynamics in Mg(BH$_4$)$_2$ were centred on the α- and β-modification [408,409], while the focus here was on the local environment of the BH$_4$ tetrahedra in the γ-polymorph and its amorphous counterpart, which both have very similar linear H$_2$BH$_2$ – Mg – H$_2$BH$_2$ chains (compare inset in Fig. 12) [215]. The different polymorphs of Mg(BH$_4$)$_2$ also exhibit distinct vibrational spectra explored by inelastic neutron scattering [410]. An overview of structure and dynamic investigations in borohydrides by neutron scattering techniques can be found from [411].

In comparison to alkali metal borohydrides, two points are noteworthy. First, in contrast to the alkali metal borohydrides, no order-disorder transition has been reported for the alkaline-earth borohydrides. Especially in LiBH$_4$, this transition is strongly connected to the enhanced Li$^+$ conductivity and especially the reorientational freedom of the [BH$_4$] units. Second, in agreement to the findings for alkali metal borohydrides, enhanced metal-ion conduction, here of the Mg$^{2+}$ cations, has been shown experimentally in Mg(BH$_4$)$_2$ based compounds, which might also be coupled to the rotational mobility of the [BH$_4$] tetrahedra. These findings encourage further research activities towards solid-states conductors for Mg-ASSBs [395,397,412,413].

# 5. Dedicated characterization tools

## 5.1. Experimental methods for thin films

### 5.1.1. Neutron Reflectometry

Neutron reflectometry is a non-destructive experimental method that utilizes the way neutrons are reflected by sufficiently large (> 100 mm$^2$) and flat surfaces to obtain structural information about the composition, roughness and thickness of thin films and other layered samples with layer thicknesses of 3 – 200 nm and with sub-nanometre resolution [414,415]. In a neutron reflectometry experiment, the sample is illuminated by a neutron beam under a small angle (θ < 5°). Subsequently, the reflectivity of the sample is measured by determining the relative amount of neutrons reflected by the sample as a function of the momentum transfer of the neutron upon reflection, $Q = \frac{4\pi}{\lambda} \sin \theta$. As $Q$ depends on both the neutrons wavelength $\lambda$ and the incident angle $\theta$, $Q$ can be varied by changing either one of these parameters. While using a monochromatic neutron beam in combination with a varying incident angle is most straightforward, measuring at a fixed angle is typically more convenient for kinetic studies. In this method, the time-of-flight method, a pulsed white neutron beam is used, and the neutrons wavelength is discriminated by utilizing the fact that the neutron's velocity is wavelength dependent. Subsequently, the measured reflectogram is fitted to a simulated one based on a structural model, yielding information about the composition (e.g. hydrogen) content, thickness and roughness of all the layers of the thin film.

Neutron reflectometry is especially a powerful technique to study thin films of metal hydrides. The reflectivity of neutrons by a material is determined by the scattering length density (SLD), $SLD = \sum_{i=i}^{n} N_i b_i$, i.e. a material property that depends on the atomic number density $N$ of isotope $i$ and the isotope-dependent scattering length $b$ (a well-known quantity). Owing to the large and negative scattering length of hydrogen ($b_H$ = -3.7 fm), neutrons are, in contrast to X-rays, sensitive to hydrogen. As neutrons are also highly penetrating, it allows for a complex sample environment and, thus, *in-situ* studying of the hydrogenation (kinetics) of thin films and is able to determine hydrogen concentration profiles in the direction normal to the film's surface.

Neutron reflectometry has for example widely been used to study magnesium-based thin films. Fritzsche *et al.* [416] and Kallisvaart *et al.* [417] used neutron reflectometry to study the hydrogenation kinetics and structural response of $Mg_{1-y}Al_y$ and other alloy thin films, determining the doping dependence of the kinetics and amount of hydrogen absorbed during hydrogenation. Dura *et al.* [418] studied palladium-capped magnesium thin films and discovered that once the film hydrogenated from Mg to $MgH_2$, the film retained its 25% increase in thickness upon desorption to Mg by incorporating voids. The formation of a porous film explains the different absorption kinetics after the initial exposure to hydrogen. Bannenberg *et al.* [419] studied a thin film of magnesium sandwiched between two titanium layers and capped with palladium. In this model system, the hydrogenation of α-$MgD_x$ to the insulating β-$MgD_{2-x}$ phase is governed by a nucleation-and-growth mechanism in which domains can grow of up to millimetres in size [420,421]. By combining optical transmission measurements with neutron reflectometry, they showed that the hydrogen solubilities of both the α-$MgD_x$ and the β-$MgD_{2-x}$ deviated considerably from the solubility limits in bulk magnesium during the phase transformation. It suggests that the enhanced kinetics of phase transformations in nanostructured systems is not only the result of reduced length scales but also caused by the enlarged solubility in the parent phases.

## 5.1.2. Nuclear Reaction Analysis

Nuclear Reaction Analysis (NRA) uses a nuclear reaction between hydrogen atoms and an incident ion beam to depth-dependent probe the hydrogen concentration of flat samples [414,422]. Typically, a flat sample is irradiated with a monochromatic beam of $^{15}$N ion that are accelerated to a specific energy. Subsequently, the ion beam may react with the $^{1}$H atoms present in the sample, forming an excited $^{16}$O nucleus that subsequently relaxes to the $^{12}$C ground state while emitting an α and γ(4.44 MeV) particle:

$$^{15}N + {}^{1}H \rightarrow {}^{16}O \rightarrow {}^{12}C + \alpha + \gamma(4.44 \text{ MeV}) \qquad (17)$$

As such, the amount of detected gamma's with this specific energy originating from the sample is thus an indirect measure of the number of hydrogen atoms present in the sample.

The probability (cross-section) that the $^{1}$H($^{15}$N,αγ)$^{12}$C reaction occurs has a strong dependence on the energy of the $^{15}$N ion, showing a narrow resonance peak around 6.385 MeV with a width of 1.8 keV. As the chance that a reaction occurs outside this energy window (off-resonance) is $10^4$ times lower, one may under most conditions neglect this probability. As such, one can probe the amount of hydrogen present on the surface of a material by accelerating the $^{15}$N beam to 6.385 MeV. In order to obtain depth-dependent concentration profiles of the sub-surface hydrogen in the material, one accelerates the beam to energies slightly exceeding the resonance energy. As the ions penetrate into the sample they experience an energy loss of about 1-4 keV/nm for most materials. Hence, by scanning the incident's beam energy, a depth-dependent hydrogen concentration can be obtained as the energy of the $^{15}$N beam matches the resonance energy at different depths. Typically, a 1-5 nm depth resolution is realised while relatively low (> 100 ppm) hydrogen concentrations can be detected for depths up to 2-4 μm. Similar to neutron reflectometry, NRA requires the samples to be flat in order to obtain a good depth resolution.

As NRA measurements are typically performed under ultra-high vacuum conditions to minimize discharges and interactions of the beam with gaseous atoms, performing measurements in which the sample is exposed *in-situ* to varying hydrogen pressures is challenging and can typically only be conducted at relatively low hydrogen pressures (< 1 mbar [423], usually much lower pressures). To this end, NRA has mainly been used to study the presence of hydrogen at surfaces, interfaces and absorbed within layers of materials that are stable under (ultra)high vacuum conditions (see Wilde and Fukutani [422] for a recent comprehensive review of applications).

## 5.1.3. Hydrogenography

Hydrogenography utilizes the fact that the optical transmission of metal hydrides changes upon hydrogenation to study the properties of thin films or arrays of nanoparticles deposited on an optically transparent substrate [424]. In a typical hydrogenography measurement, the optical transmission as a function of time is monitored at a set temperature while the (partial) hydrogen pressure can be varied stepwise. As such, pressure-optical transmission-isotherms (PTIs) can be obtained. Similar to measure pressure-composition-isotherms, these optical transmission measurements allow the detection of hysteresis and the determination of the enthalpy and entropy of formation through Van 't Hoff's equation:

$$\ln \frac{P_{eq}}{P_0} = \frac{\Delta H}{RT} - \frac{\Delta S_0}{R}, \qquad (18)$$

where $P_{eq}$ is the equilibrium pressure, $P_0$ the standard pressure, $R$ the gas constant, $T$ the absolute temperature and $\Delta H$ and $\Delta S_0$ are the enthalpy (kJ mol$^{-1}$ H$_2$) and entropy (JK$^{-1}$ mol$^{-1}$ H$_2$) of the hydrogenation reaction, respectively. Indeed, by measuring the optical transmission of a thin film metal hydride as a function of the hydrogen pressure for different temperatures, the temperature dependence of $P_{eq}$ can be determined at a given optical transmission (i.e. halfway the plateau). Subsequently, from plotting $P_{eq}$ versus the inverse of temperature, $\Delta H/R$ (slope) and -$\Delta S_0/R$ (intercept) can be obtained. In this analysis, it is assumed that the experimental conditions (temperature, pressure) at which the optical transmission is the same, correspond to the same hydrogen-to-metal ratio of the metal hydride layer.

Hydrogenography is not confined to materials showing a metal to insulating transition. Indeed, metals such as titanium, hafnium and palladium can also be studied as long as their layer thickness is sufficiently thin (< 100 nm). In addition, it is important to realize that the thermodynamics of thin films may deviate from bulk materials owing to clamping of the film to the support as well as from the increased surface-to-volume ratios.

Generally speaking, a detailed understanding of the changes induced by the hydrogenation of the material on the optical transmission is required to relate the measured changes in the optical transmission to the difference in the hydrogen-to-metal ratio. As such, additional measurements as e.g. neutron reflectometry measurements need to be performed to translate the changes in optical transmission to the hydrogen-to-metal ratio of the sample. In practice, however, it turns out that the natural logarithm of the relative transmission with respect to the reference (unloaded) state, i.e. *Ln(T/T$_{ref}$)*, is often proportional to the hydrogen-to-metal ratio. While it directly follows from Lambert-Beer's Law that such a relation is expected for a two phase system with varying fractions of the two phases [24], this relation has also been observed experimentally for solid solutions [24,332-334,425,426]. It suggests that the effect of hydrogen sorption on the optical properties of the metal hydride is independent of the hydrogen concentration.

Hydrogenography has some distinct advantages over more conventional methods to study the properties of metal hydrides. The enhanced hydrogen (de)sorption kinetics of thin films over bulk materials allows for much faster measurements. In addition, hydrogenography facilitates the simultaneous measurement of a large number of samples, either by multiple samples with dimensions of typically 10 x 10 mm$^2$ at the same time, or by examining large area films with compositional gradients. These gradients can for instance be obtained by co-deposition from two, three or more off-centred metal sources in a magnetron sputtering set-up [427,428].

## 5.2. Knudsen effusion method for gaseous phase of hydrides

The development of suitable solid-state hydrogen storage and ion conducting hydride materials must pass through the characterization and thermodynamic evaluation of the specific gas phase, being present above a condensed phase, and taking in consideration the following experimental facts:

- When the quantification of partial pressures is possible, the range of low partial pressures corresponds to an ideal gas solution, i.e. the compositions and the partial pressures are identical; this is often the case up to at least 1 Pa pressure and far from the critical point;

- The quantities of material involved in the evaporation flows required for a gas phase measurement are much smaller than those that can exist in the condensed phase, and the disturbances caused by the measurement can therefore be limited;

- The identification of the composition of a gaseous phase of multi-component systems using a pressure determination method with adjacent analytical analyzer (e.g. mass spectrometry, gas chromatography, etc.) can give the information of adsorption/desorption mechanisms.

The Knudsen effusion method [429-431], in combination for instance with a mass spectrometer, allows the measurement of the vapor pressure in the gaseous phase between $10^{-6}$ - 10 Pa until 600 K. In the simplest thermodynamic application of the Knudsen effusion method, the vapor, composed of a single species, is in equilibrium with its congruently evaporating condensed phase, and flows from the sample surface to the isothermal container and then through a small orifice into an evacuated space at high vacuum. Since the vapor has a mean free path that is longer than the orifice diameter, the effusing molecules behave like a well-defined molecular beam of atoms or molecules moving in nearly collision-free trajectories with particles distribution that are easily calculable from gas kinetic theory. The Knudsen equation relates the total vapor pressure $p$ above the sample to the mass effusion rate by:

$$p = \frac{\Delta m \sqrt{2\pi RT}}{sC\sqrt{M}\Delta t} \qquad (19)$$

where $\Delta m$ is the mass loss during the time $\Delta t$ at the temperature $T$ of the experiment, R the gas constant, M the mass of a molecule in the vapor, s the orifice area and C a correction coefficient, the Clausing factor, [432] that quantitatively predicted the departure from the ideal cosine distribution (thin edge orifices) for beams effusing from non-ideal orifices (usually cylindrical or conical orifices).

Coupling the Knudsen effusion method with mass spectrometry [433-435] has the attractive features of high sensitivity and resolution under high vacuum conditions, that is especially a useful tool for the qualitative and quantitative detection of gaseous species. By heating or by milling a sample (solid, liquid, or gaseous) in the Knudsen cell, a molecular beam is formed. This effused flow is ionized, for example by bombarding it with electrons into an ionization chamber. This may cause some of the sample's molecules to breaks into charged fragments or simply become charged without fragmenting. Then, all ions formed by different ionization processes are accelerated by an electric field and then they are separated by a magnetic and/or an electric field according to their mass-to-charge ratio. A Faraday cup or a secondary electron multiplier is used for ion current measurements. The temperature of the sample in the Knudsen cell is measured either with an optical pyrometer or with a thermocouple (Fig. 13a). Knudsen cell temperatures and intensities of ion currents related to gaseous species are the quantities measured in the course of an investigation by Knudsen effusion mass spectrometry.

Partial pressures of species $i$, $p_i$, at temperature $T$ are defined by the basic mass spectrometric relation [436]:

$$p_i \cdot S_i = I_{(i,j)} \cdot T \tag{20}$$

where $I_{(j,i)}$ is the sum of intensities of the ion currents $j$ originating from molecule $i$, where $S_i$ is the sensitivity of the apparatus and $T$ the temperature of the neutral species (i.e. of the Knudsen cell at the time of its evaporation). Partial pressures are obtained by calibration that is generally performed for each measurement separately in order to determine the sensitivity $S_i$ that includes several factors as:

$$S_i = G \cdot \eta_{i,j} \cdot \sigma_i(E) \cdot \gamma_{i,j} \cdot f_{i,j} \tag{21}$$

where $G$ is a geometric factor involving the solid angle between the Knudsen orifice and the source aperture defining the useful molecular beam, $\eta_{i,j}$ is the ion transmission factor in the spectrometric analyser, $\sigma_i(E)$ is the total ionisation cross section of the molecule $i$ at $E$ ionization energy, $\gamma_{i,j}$ is the efficiency of the detector and $f_{i,j}$ is the isotopic abundance known or calculated from the constituent atoms of the ion.

There are only a very limited number of studies on the hydrides gaseous phase owing to complexities of the experimental set-up. From metal hydrides, various studies about the gaseous phase showed some aspects:

- Quantitative determination of gaseous species in the binary systems as H-M (Li,B,…) showed that $H_2(g)$ is the major one, followed by gaseous hydrides such as MH(g), $MH_2(g)$, or polymers such as $M_2H_2(g)$. The relative importance of these gaseous hydrides depends on the composition of the condensed phase: for hydrogen rich composition side of a metal hydride it is $H_2(g)$ which predominates and on the M side are the metal gaseous hydrides;

- When varying the composition, especially for the M rich side, some liquid phase may appear (this is the case in Li-H system for Li(l)-LiH(s) side) [437];

- Determination of the composition of the gas phase by mass spectrometry may present difficulties especially for the measurement of $H_2(g)$, i.e. a so-called permanent gas that is subject to multiple reflections between the two compartments furnace and source, and even in the ionization chamber. It is then recommended to use a restricted collimation [438] equipped with an adequate molecular beam shutter [439,440] and an *in situ* liquid nitrogen cooled ionization chamber [441];

- Materials for building the effusion cell must be chemically compatible with the condensed hydride, but must also remain impervious (especially with temperature) to hydrogen (avoid diffusion through the walls) so as not to disturb the measurements;

- Handling the sample and loading into the effusion cell and then the cell into the mass spectrometer must have all adapting facilities to carry the sample under air inert conditions.

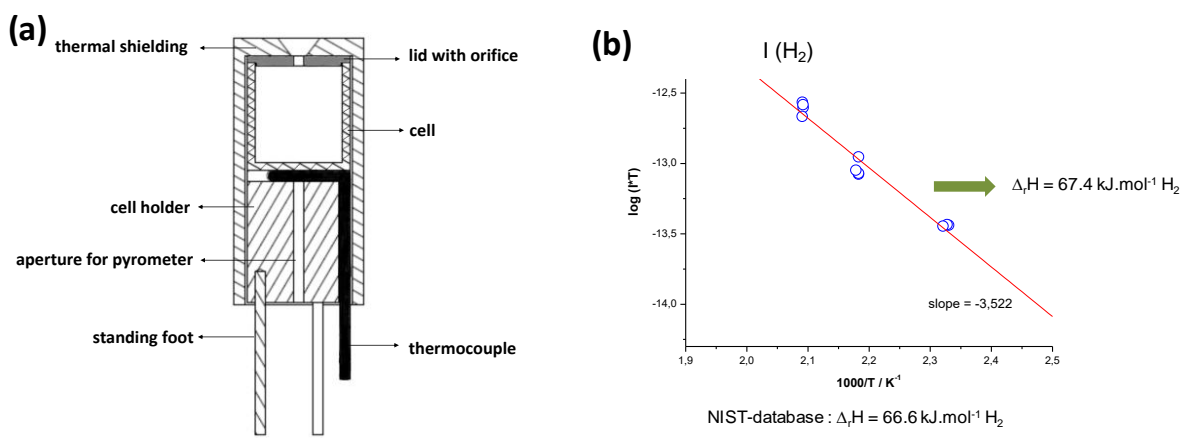

**Figure 13.**
(a) Knudsen cell schema and (b) logarithmic variation of the product $I*T$ (proportional to pressure) for $H_2$ as a function of the inverse of the temperature for $LiBH_4$.

In 1966, Baylis *et al*. [442] used a mass spectrometer equipped with a Knudsen cell, introducing diborane ($B_2H_6(g)$) at low pressure to observe its pyrolysis. From the spectral analysis, the borane ($BH_3$) spectrum was identified at temperatures above 650 K as well as the production of $H_2(g)$ molecule associated with a dark deposit (probably Li and B) in the Knudsen cell.

For more complex systems as $H-M^I(Li,…)-M^{III}$ (B,…), Züttel *et al*. [443] observed the desorption peak of $H_2(g)$ during temperature ramps up to 100 °C in $LiBH_4$ using a mass spectrometer in an ultra-high vacuum (UHV) chamber that is associated with structural transformation of $LiBH_4$. Indeed, temperature programmed desorption method (i.e. very slow $T$ ramp type as 5 °C min$^{-1}$) can be used to monitor the different steps of desorption and to give a mechanism of evaporation kinetics. Particularly, the peaks (which in fact more resemble bumps-like) observed as a function of temperature in the evolution of the $H_2^+$ ion (it could be the same for $MH^+$) correspond to different desorption energies evaluated from the bumps summit. These energies are H-solid breaking bonds and are not the enthalpies of formation for gaseous species. These studies are different from the gas phase study (aimed to acquire thermodynamic data about gaseous species) performed with the Knudsen method, although the latter method may also be used for a desorption kinetics experiment.

An understanding of the processes that occur upon heating $LiBH_4$ is important for its practical use. Zhinzhin *et al*. [444] studied the influence of mechanical activation on the thermal decomposition of $LiBH_4$. They observed that the mechanical activation noticeably enhances the reactivity of lithium tetrahydroborate. Two mass spectrometric studies [445,446] have been reported that monitor the gaseous species during $LiBH_4$ decomposition/vaporization where $H_2$ is the most important species before and after the mechanical treatment. Gas evolution from a treated sample starts at lower temperature and strongly increases at the second melting step. However, Zhinzhin *et al*. [444] notices that while the temperature maxima of gas desorption are the same for both studies, in [445] there is a hydrogen excess, whereas in [446] measurements are performed during continuous pumping. Using an airtight Knudsen cell made of stainless steel with controlled opening [447] coupled with a mass spectrometer, El Kharbachi [448] measured the total vapor pressure of $LiBH_4$ solid phase and reported its variation as a function of the inverse temperature from 373 to 573 K (Fig. 13b). The derived enthalpy of reaction, considering the main equilibrium $LiBH_4 \leftrightarrow LiH + B + 2H_2$, is in agreement with the compiled value in NIST-databased, the latter being based on the experiments of Davis *et al.* [449], using bomb calorimetry in 1949.

# 6. Summary and perspectives

**Solid-state hydrogen storage.** As hydrogen gas at ambient conditions has very low volumetric energy density, effective storage of hydrogen is crucial for its implementation as an energy carrier for stationary and mobile applications. One of the methods of hydrogen storage being developed for the last few decades utilizes hydrogen-rich chemical compounds as solid-state stores. Both simple and complex hydrides are potentially interesting basic materials for this purpose, however, the latter are able to fulfill the required gravimetric and volumetric hydrogen capacity for much broader range of chemical composition. As thermodynamic and kinetic factors limit the performance of chemical hydrogen stores, continuous efforts are still required to obtain cost-effective materials with high storage efficiency. Although advances have been made in the last decades, the prospective advantages of hydrogen storage in metal hydrides faded partly away owing to the rapid development of alternative technologies as Li-ion batteries and gaseous hydrogen storage in lightweight high-pressure vessels. However, the potential materials for solid-state hydrogen storage are still under development, and various metal hydrides may be unbeaten in some aspects of hydrogen processing. The palladium-based purifying membranes are a classic example here.

The Ti-V based alloys have been known for decades, and many studies have been conducted. The thermodynamic properties of the Ti-V hydrides can be tuned by changing the atomic ratio of the Ti/V and the addition of other alloying elements such as Fe, Cr, and Mn. The hydrogen sorption kinetics is known to be fast, especially when the alloy is nanostructurally prepared. The hydrogen desorption proceeds at least three steps involving phase-structural changes depending on the alloying elements. However, desorption temperature and hydrogen capacity are the trade-off variables because the addition of alloying elements, in general, decreases the hydrogen capacity of the Ti-V based hydrides. Since the Ti-V alloys can be tuned in rather large ranges, the application of the Ti-V alloys ranges from low temperatures such as for stationary hydrogen storage or medium temperature such as in hydrogen separation processing. The challenge for further development will be optimizing the hydrogen storage capacity while meeting the desorption temperature.

Intermetallic compounds, as TiFe, present renewed interests owing to their low cost and highly tunable hydrogenation properties as concerned equilibrium pressure and temperature close to room temperature, well adapted to electrolyzer output pressures. The productive literature on TiFe H-sorption properties and their possible adjustment by elemental substitutions make a consolidated scientific base for moving toward real applications and prototype systems, especially for stationary energy storage. However, though excellent volumetric capacity, the moderate gravimetric hydrogen capacity of these metal hydrides make their use for mobile application difficult. Nevertheless, the absence of critical raw material in their formulation, their fast kinetics and good cycling properties make them a perfect choice for integrated systems in the production, storage and use of green hydrogen and for energy balancing in smart grids. Future efforts should be focused on the collection of hydrogenation properties and characterization of these materials in scaled-up systems, to move forward the exploitation of TiFe-based compounds in real applications and into the market, implementing and diffusing the hydrogen economy locally and globally.

HEA is the new classification for alloys usually containing five or more elements owning interesting mechanical properties. In recent years, these alloys have demonstrated promising properties towards hydrogen absorption as solid-state storing materials. With only a dozen or more of reported articles, some *bcc* HEAs have demonstrated superior hydrogen uptake as compared to conventional *bcc* alloys. Additionally, it has been reported that HEA can rapidly absorb hydrogen at room temperature conditions, reaching full capacity within minutes at low equilibrium pressures. Another interesting feature is the single-step reaction of the hydrogenation process, *bcc* ↔ (pseudo)*fcc*. However, the main drawback is the desorption that needs high temperature and vacuum. The main characteristic(s) that grant HEA these properties remains unclear and more research is needed in the future to fully understand these systems and propose a material design strategy able to respond to all criteria for use in a practical storage device.

As metal borohydrides contain relatively the largest amount of hydrogen, a remarkable expansion of their chemistry has been observed recently. This was possible due to significant developments in synthesis of new compounds *via* mechanochemical and solvent-mediated routes of general applicability. Such methods can be utilized for preparation of broad range of products and allow for fast screening of hydrogen storage properties of numerous borohydrides based on main-group and transition metals. Besides hydrogen storage, borohydrides have been tested as potential luminescent or magnetic materials, and as source of potentially useful decomposition products. The latter includes $LiZn_2(BH_4)_5$ utilized as a convenient, small-scale source of diborane, or various borohydrides tested as precursors for refractory borides.

**Optical sensors.** Metal hydride based hydrogen sensors are considered to be an effective way to accurately sense hydrogen. Although many different designs exist, all these sensors have in common that they utilize the propensity of metal hydrides to hydrogenate when exposed to a partial hydrogen pressure, which, in turn, results in volumetric expansion of the metal hydride and a change in its optical and electronic properties. By probing one of these properties, the hydrogen pressure in the environment of the metal hydride sensors can thus be determined. Although resistivity based hydrogen films are already considered for long time, hydrogen sensors with an optical readout, based for example on optical transmission/reflectivity or LSPR, have the particular advantage that no currents are required in the sensing area, thus eliminating a potential safety hazard.

Palladium has been the most frequently used material in metal-hydride based sensors owing to its large contrast, its modest sensing range and its propensity to dissociate molecular hydrogen at room temperature. However, the first-order metal-to-metal hydride phase transition in palladium has the distinct disadvantage that the sensor output is highly hysteretic. To this end, various other materials, including palladium alloys and transition metals have been considered to obtain sensing response free of hysteresis. Another challenge is to improve the resistance to chemical species other than hydrogen of the sensors, being key for the large-scale implementation of the metal-hydride based sensors.

**Electrochemical energy storage.** Conversion-type metal hydride anodes remain as potential candidates for LIBs, owing to their high capacity ($MgH_2$ ~2000 mAh $g^{-1}$) compared to the commercialized graphite (~372 mAh $g^{-1}$). Despite the efforts made to improve the long-term cyclability of $MgH_2$ anode, the reversibility of the composite Mg/2LiH is still a challenge. The capacity fading is more pronounced when a carbonate-liquid electrolyte is used. However, this electrode system shows a real potential in solid-state battery with light-weight $LiBH_4$-based electrolytes [450,451]. Owing to the high capacity of MgH$_2$ and the beneficial aspects of ASSBs, it is expected that research would consider further these systems in order to reach high energy density electrochemical storage prototypes [452]. Some computational techniques such as quantum DFT and AIMD are helping the acceleration of research and development of better metal hydrides for electrochemical electrode materials especially combining with the cutting-edge materials genome methods such as high-throughput computing, data mining and so on.

The halide substitution in LiBH$_4$ has been a breakthrough in suppressing the phase transition and maintaining the properties of the high temperature phase at room temperature, i.e. ionic conductivity. This discovery has been followed by the study of interface issues in lithium cells, so that to ensure cycling stability of the cells; however, these cells need to be operated at high temperatures (50-100 °C), which may speed up the appearance of surface degradation aspects. Working below 50 °C became a necessity and a key factor to target larger application fields. A series of materials based on LiBH$_4$-LiX-P$_2$S$_5$ (X = halogen) have promising ionic conductivity (~$10^{-3}$ S cm$^{-1}$) and chemical properties for lithium batteries. Their different treatments mostly end up with a dominant amorphous structure due to the presence of P$_2$S$_5$. Efforts are needed to clarify their structure and phases' stability, and how the change in the structural properties will affect ions conduction and many other relevant properties for batteries (mechanical, chemical, thermal etc).

On the other hand, ASSBs based on magnesium are still a very young scientific topic and have exciting times ahead. Many challenges, especially technical challenges that are related to the design of low cost materials with high ionic conductivities and, at the same time, low environmental impact, still have to be considered. Mg(BH$_4$)$_2$ and solvated derivatives, such as ethylenediamine, diglyme and ammonia, tend to form stable compounds. Conductivities were reported such as σ = 6·10$^{-5}$ S·cm$^{-1}$ at 343 K of Mg(en)$_1$(BH$_4$)$_2$ [397], σ = 2·10$^{-5}$ S·cm$^{-1}$ at 350 K of Mg(BH$_4$)$_2$-diglyme$_{0.5}$ [398] and σ = 3.3·10$^{-4}$ S cm$^{-1}$ at T = 353 K for Mg(BH$_4$)$_2$·NH$_3$ [401]. The later study even postulated a new approach for the Mg diffusion with dihydrogen bonds being involved [401].

In most of the aforementioned materials amorphous Mg(BH$_4$)$_2$ was formed as a byproduct [397,398]. Its influence on the conduction properties was reviewed here, as it was postulated that the amorphous phase is aiding to increase Mg-ion conductivity. EIS data stated that the conductivity of the amorphous phase is indeed ~2 orders of magnitude higher than the as-received γ-Mg(BH$_4$)$_2$ at 353 K. From PDF analysis and the reported similar local building blocks, it was suggested that also similar conduction pathways are at play. QENS data indicated a higher fraction of activated rotations in the amorphous sample. Thus it is assumed that the conduction process in amorphous Mg(BH$_4$)$_2$ is supported by rotating [BH$_4$] units. Upon crystallization at 373 K, the number of rotations decreases as well as the conductivity does. In general, Mg(BH$_4$)$_2$ with all its derivatives will make an important contribution to future Mg-based ASSBs, while its amorphous structure should not be neglected in those investigations [215].

**Characterization tools.** Several specialized characterizations tools are at hand to researchers studying thin film and bulk metal hydrides. For thin films, although neutron reflectometry and NRA can both quantify the hydrogen concentrations, neutron reflectometry has the particular advantage that such experiments do not have to be performed under high vacuum conditions. As such, while NRA can typically achieve a higher lateral resolution and detect smaller quantities of hydrogen, neutron reflectometry facilitates the simultaneous determination of the out-of-plane expansion and the hydrogen concentration in thin films in combination with complex sample environment, thus i.e. in the presence of a hydrogen atmosphere. In addition, pressure-transmission-isotherms can be measured with the dedicated technique of hydrogenography. In such experiments, the optical transmission of thin film metal hydrides deposited on a transparent substrate are measured, allowing e.g. the detection of hysteresis and the determination of the enthalpy and entropy of formation through Van 't Hoff's equation. Hydrogenography has the particular advantage over other methods that the fast (de)sorption kinetics of thin films over bulk materials allows for much faster measurements and that many different samples can be measured simultaneously.

As a method of choice for the gaseous phase of hydride-based materials, the Knudsen effusion method coupled with mass spectrometry is a useful tool for both qualitative and quantitative detection of gaseous species, in addition to the study of the kinetic processes and adsorption/desorption mechanisms of solid state materials and thermodynamic properties (enthalpy of formation, Gibbs free energy, …), in order to contribute to model phase diagrams of complex systems such as metal borohydrides and their mixtures which are still lacking in literature [309,453-455].

# Acknowledgements

TJ's contribution received support from Polish National Science Centre (UMO-2016/21/D/ST5/01966). MW, WW, AS, KF and WG acknowledge the support from the HYDRA project of the Polish National Science Center, NCN (2014/15/B/ST5/05012). PO's work has been financed from the project "Diamond Grant" of the Polish Ministry of Science and Higher Education (DI2015 008845, contract No. 322268). The research was carried out with the use of CePT infrastructure financed by the European Union – the European Regional Development Fund within the Operational Programme "Innovative economy" for 2007–2013 (POIG.02.02.00-14-024/08-00). ZQ's computational contribution received support from the Natural Science Foundation of China (51801113).

EMD, FC, ML and MB acknowledges support through the HyCARE project. This project has received funding from the Fuel Cells and Hydrogen 2 Joint Undertaking (JU) under grant agreement No 826352. The JU receives support from the European Union's Horizon 2020 research and innovation programme and Hydrogen Europe and Hydrogen Europe Research.

This work contributes to the research performed at CELEST (Center for Electrochemical Energy Storage Ulm-Karlsruhe).60